\documentclass[11pt]{article}
\usepackage{amsfonts}
\usepackage{amssymb}
\usepackage{latexsym}
\usepackage{graphicx}
\usepackage[english]{babel}
\begin{document}
\input epsf

\renewcommand{\theequation}{\arabic{section}.\arabic{equation}}
\def\p{\partial}
\def\h{{1\over 2}}
\def\be{\begin{equation}}
\def\bea{\begin{eqnarray}}
\def\ee{\end{equation}}
\def\eea{\end{eqnarray}}
\def\d{\partial}
\def\la{\lambda}
\def\eps{\epsilon}
\def\bb{\bigskip}
\def\mm{\medskip}
\newcommand{\dm}{\begin{displaymath}}
\newcommand{\edm}{\end{displaymath}}
\renewcommand{\b}{\tilde{B}}
\newcommand{\gm}{\Gamma}
\newcommand{\ac}[2]{\ensuremath{\{ #1, #2 \}}}
\renewcommand{\ell}{l}
\newcommand{\z}{\ell}
\newcommand{\newsection}[1]{\section{#1} \setcounter{equation}{0}}
\def\bb{$\bullet$}
\def\Qbar{{\bar Q}_1}
\def\QPbar{{\bar Q}_p}

\def\q{\quad}

\def\bn{B_\circ}

\let\a=\alpha \let\b=\beta \let\g=\gamma \let\d=\delta \let\e=\epsilon
\let\c=\chi \let\th=\theta  \let\k=\kappa
\let\l=\lambda \let\m=\mu \let\n=\nu \let\x=\xi \let\r=\rho
\let\s=\sigma \let\t=\tau
\let\vp=\varphi \let\vep=\varepsilon
\let\w=\omega      \let\G=\Gamma \let\D=\Delta \let\Th=\Theta
                     \let\P=\Pi \let\S=\Sigma

\def\h{{1\over 2}}
\def\t{\tilde}
\def\r{\rightarrow}
\def\nn{\nonumber\\}
\let\bm=\bibitem
\def\Kt{{\tilde K}}

\let\p=\partial

\begin{flushright}
\end{flushright}
\vspace{20mm}
\begin{center}
{\LARGE  Dynamics of supertubes}
\\
\vspace{18mm}
{\bf   Stefano Giusto\footnote{giusto@mps.ohio-state.edu}, Samir D. Mathur\footnote{mathur@mps.ohio-state.edu} and Yogesh K. Srivastava\footnote{yogesh@mps.ohio-state.edu}}\\

\vspace{8mm}
Department of Physics,\\ The Ohio State University,\\ Columbus,
OH 43210, USA\\
\vspace{4mm}
\end{center}
\vspace{10mm}
\thispagestyle{empty}
\begin{abstract}

We find the evolution of arbitrary excitations on 2-charge supertubes, by mapping the supertube to a string carrying traveling waves. We argue that when the coupling is increased from zero the energy of excitation leaks off to infinity, and when the coupling is increased still further a new set of long lived excitations emerge. We relate the excitations at small and large couplings to excitations in two  different phases in the dual CFT. We conjecture  a way to distinguish bound  states from unbound states among 3-charge BPS geometries; this would identify black hole microstates among the complete set of BPS geometries.

\end{abstract}
\newpage
\setcounter{page}{1}
\renewcommand{\theequation}{\arabic{section}.\arabic{equation}}
\section{Introduction}\label{intr}
\setcounter{equation}{0}

A supersymmetric brane in Type II string theory is a $1/2$ BPS object. The bound state of $N$ identical branes (wrapped on a torus) can be mapped by duality to a massless quantum with momentum $P=N/R$ on a circle of radius $R$. Thus the bound state has degeneracy 256, regardless of $N$.

The situation is very different for $1/4$ BPS states. Such states can be made in many duality related ways: NS1-P, NS1-D0, D0-D4, D1-D5 etc. If the two charges are $n_1, n_2$ then the degeneracy of the bound state is $Exp[2\pi\sqrt{2}\sqrt{n_1n_2}]$ \cite{vafa,sen}. In the classical limit $n_1, n_2\r \infty$ this degeneracy manifests itself as a continuous family of solutions. Examples are the 2-charge D1-D5 solutions found in \cite{lm4} and the supertubes constructed in \cite{supertube,baka,kmpw}. These 2-charge states are important because they give the simplest example of a black hole type entropy \cite{sen}.

In this paper we address the question: What is the low energy dynamics of such $1/4$ BPS states? We will perform some calculations to arrive at a conjecture for the answer. The behavior of the system can depend on whether the coupling is small or large, and whether we have bound states or unbound states. For this reason we first give an overview of  {\it possible}
dynamical behaviors, and then summarize our computations and conclusions.

\subsection{Possibilities for low energy dynamics}\label{poss}

In the following it will be assumed that all compactifications are toroidal, and all branes are wrapped on these compact directions in a way that  preserves $1/4$  supersymmetry.

\medskip

(a) {\it `Drift' on moduli space:}\quad A D0 brane can be placed at rest near a D4 brane; there is no force between the branes. Thus we have a moduli space of possibilities for the relative separation. If we give the branes a small relative velocity $v$ then we get a $\sim v^2$ force, and the resulting motion can be described by `motion on moduli space' \cite{dkps}. More generally, we can make 3-charge black holes that are $1/8$ BPS and their slow motion will be described by motion on a moduli space \cite{kami,mist}. 
For later use we make the required limits explicit: The velocity $v$ is  $O(\epsilon)$, the time over which we follow the motion is $O(1/\epsilon)$, and the distance in moduli space over which the configuration `drifts' is $O(1)$
\be
v\sim \epsilon, ~~~\Delta t\sim {1\over \epsilon}, ~~~\Delta x\sim 1, ~~~~~~~(\epsilon\r 0)
\label{drift}
\ee
Note that the different branes or black holes involved here are not bound to each other.

\medskip

(b) {\it Oscillations:}\quad Consider  branes carrying just one  charge, and let these be NS1 for concreteness.   Then the force between branes is $\sim v^4$. So for the relative motion between such branes we again get `drift' on moduli space except that the moduli space is flat \cite{khuri}. But we can also focus on just one brane and study its low energy excitations. These will be vibration modes along the brane, with the amplitude for each harmonic behaving like a harmonic oscillator. Calling the amplitude for a given harmonic $A_n\equiv x$ we note that $x$ will  have the time evolution $x= \bar x\cos(\omega t+\phi)$. Setting  $\bar x=\epsilon$ for a small deformation, the analogue of (\ref{drift}) is
\be
v\sim \epsilon, ~~~\Delta t\sim 1, ~~~\Delta x\sim \epsilon, ~~~~~~~(\epsilon\r 0)
\label{osc}
\ee
where we have assumed that we are not looking at a zero mode $\omega=0$. For the zero mode we will have the behavior
\be
x=x_0+vt
\label{zero}
\ee
and we get `drift' over configuration space with characteristics given  by (\ref{drift}).

\medskip

(c) {\it `Quasi-oscillations':} \quad Consider a charged particle  free to move in the $x-y$ plane in a uniform magnetic field $F_{xy}=B$. The particle can be placed at rest at any position on the plane, and it has the same energy at all these points. Thus far its behavior  looks like that of a system with a zero mode. But if we give the particle a small velocity then it describes a small circle near its original position, instead of `drifting' along the plane. The motion is described by
\be
v\sim \epsilon, ~~~\Delta t\sim 1, ~~~\Delta x\sim \epsilon ~~~~~~~(\epsilon\r 0)
\label{quasi}
\ee
Thus even though we may have a continuous family of energetically degenerate configurations, this does not mean that the dynamics will be a `drift' along this space.

\medskip

(d) {\it Gravitational radiation:} \quad We are going to give our system a small energy above extremality. But the system is coupled to Type II supergravity, and there are massless quanta in this theory. Thus any energy we place on our branes can leave the branes and become radiation flowing off to infinity. There will of course always be some radiation from any motion in the system,  but the relevant issue here is the time scale over which energy is lost to radiation. If the time scale relevant to the dynamics is $\Delta t$ then as $\epsilon\r 0$ we have to ask what fraction of the energy is lost to radiation in time $\Delta t$. If the fraction is $O(1)$, then the system is strongly coupled to the radiation field and cannot be studied by itself while ignoring the radiation. If on the other hand the fraction of energy lost to radiation goes to zero as $\epsilon\r 0$ then the radiation field decouples and radiation can be ignored in the dynamics.

\medskip

(e) {\it Excitations trapped near the brane:} \quad In the D1-D5 system we can take a limit of parameters such that the geometry has a deep `throat' region. In \cite{lm3,lm4} it was found that excitations of the supergravity field can be trapped 
for long times in this throat; equivalently, we can make standing waves that leak energy only slowly to the radiation modes outside the throat \cite{lm6}. These are oscillation modes of the supergravity fields and thus could have been listed under (b) above. We list them separately to emphasize that the fields excited need not be the ones making the original brane state; thus the excitation is not in general a collective mode of the initial fields.

\subsection{Results and conjectures}

Consider first the D1-D5 bound state geometries found in \cite{lm4}. These geometries are flat space at infinity, they have a locally $AdS_3\times S^3\times T^4$ `throat', and this throat ends smoothly in a `cap'. The geometry of the cap changes from configuration to configuration, and is parametrized by a function $\vec F(v)$. All the geometries have the same mass and charges, are $1/4$ BPS, and yield (upon quantization) different bound states of the D1-D5 system. 

What happens if we take one of these geometries and add a small energy? The bound state of D1-D5 branes has a nontrivial transverse size, so one may say that the brane charges have separated away from each other in forming the bound state. If the charges indeed behave like separate charges then we would expect  `drift on moduli space' dynamics, (type (a) in our list).  Or does the bound  state fragment into a few unbound states, which then drift away from each other? This is in principle possible, since the D1-D5 system is threshold bound. Do we stay within the class of bound geometries of \cite{lm4} but have `drift on moduli space' (\ref{drift}) between different bound state configurations (i.e. drift on the space $\vec F(v)$)? Or do we have one of the other possibilities (b)-(e)?

Now consider the opposite limit of coupling: Take a supertube in flat space. The supertube carries NS1-D0 charges, and develops a D2 `dipole' charge. This D2 brane can take on a family of possible profiles in space, giving a continuous family of $1/4$ BPS configurations. What happens if we take a supertube in any given configuration and add a small amount of energy? Is there a `drift' among the family of allowed configurations, or some other kind of behavior?

In \cite{pm} the `round supertube' was considered, and the low energy behavior yielded excitations   with time dependence $\sim e^{-i\omega t}$. Can we conclude that there is no `drift' among supertube configurations? Any drift can occur only between states that have the same values of conserved quantities. The round supertube has the maximal possible angular momentum $J$ for its charges, and is the only configuration with this $J$. So `drifting' is not an allowed behavior if we give a small excitation to this particular supertube, and we must look at the generic supertube to know if periodic behavior is the norm.

\medskip

We now list our computations and results:

\medskip

(i) First we consider the 2-charge systems in flat space (i.e., we set $g=0$). It turns out that the simplest system to analyze is NS1-P, which is given by a NS1 string wrapped $n_1$ times around a circle $S^1$, carrying $n_p$ units of momentum along the $S^1$. The added excitation creates further vibrations on the NS1. But this is just a state of the free string, and can be 
exactly solved (the classical solution is all we need for our purpose). Taking the limit $n_1, n_p\r\infty$ we extract the dynamical behavior of the supertube formed by NS1-P charges. In this way we get not only the small perturbations but arbitrary excitations of the supertube.

We then dualize from NS1-P to D0-NS1 which gives us the traditional supertube. This supertube can be described by a DBI action of a D2 brane carrying fluxes. We verify that the solution found through the NS1-P system solves the dynamical equations for the D2, both at the linear perturbation level and at the nonlinear level.

Even before doing the calculation it is easy to see that there is no `drift' over configurations in the NS1-P dynamics.  The BPS string carries a right moving wave, and the excitation just adds a left moving perturbation. Since right and left movers can be separated, the perturbation travels around the string and the string returns to its initial configuration after a time $\Delta t \sim 1$. But this behavior of the `supertube' is not an oscillation of type (b); rather it turns out to be a `quasi-oscillation' of type (c). This can be seen from the fact that even though we move the initial tube configuration towards another configuration of the same energy, the resulting motion is periodic rather than a `drift' which would result from a zero mode (\ref{zero}).

\medskip

(ii) Our goal is to move towards larger values of the coupling $g$. At $g=0$ the gravitational effect of the supertube does not manifest itself at any distance from the  supertube. Now imagine increasing $g$, till the gravitational field is significant over distances $\sim  Q$ from the supertube. Let the radius of the curve describing the supertube profile be $\sim a$. We focus on the domain
\be
 Q \ll a
\ee
Then we can look at a small segment of the supertube which looks like a straight line. But this segment is described by a geometry, and we look for small perturbations of the geometry. We solve the linearized supergravity equations around this `straight line supertube' and note that the resulting periods of the solutions agree with (i) above. 

We note however that far from the supertube the gravity solution will be a perturbation on free space with some frequency satisfying $\omega^2>0$. The only such solutions are traveling waves. For small $Q/a$ we find that  the amplitude of the solution when it reaches the approximately flat part of spacetime is small. Thus we expect that the radiation into modes of type (d) will be suppressed by a power of $ Q/a$. 

\medskip

(iii) Now imagine increasing $g$ to the point where
\be
 Q\sim a
\ee
In this situation we see no reason why the part of  the wavefunction leaking into the radiation zone should be suppressed. Thus we expect that the excitation will not be confined to the vicinity of the branes, but will be a gravitational wave that will flow off to infinity over a time of order the crossing time across the diameter of the supertube.

\medskip

(iv) We increase the coupling further so that
\be
 Q \gg a
\ee
Now the geometry has a deep `throat' and as mentioned above we find excitations which stay trapped in this throat for long times, with only a slow leakage to radiation at infinity. What is the relation between these excitations and those found in
(ii)? We argue that these two kinds of excitations are different, and represent the excitations in two different phases of the 2-charge system. These two phases were identified by looking at microscopic degrees of freedom as a function of $g$ in \cite{emission,review}, and what we see here appears to be a gravity manifestation of the transition.

\medskip

(v) All the above computations were for {\it bound states} of the 2-charge system. But we have seen above that if have
{\it unbound} states -- two different 2-charge black holes for example -- then we get `drift' modes of type (a). It looks reasonable to assume that in any coupling domain if we have two or more different bound states then the relative motion of these components will be a `drift'. For example at $g=0$ we can have two supertubes that will move at constant velocity past each other. 

It is intriguing to conjecture that this represents a basic difference between bound and unbound states: Bound states have no `drift' modes and unbound states do have one or more such modes. The importance of this conjecture is that 3-charge systems are very similar to 2-charge ones, so we would extend the conjecture to the 3-charge case as well. While all bound states can be explicitly constructed for the 2-charge case, we only know a few 3-charge bound states \cite{3charge}. There is a way to construct {\it all} 3-charge supersymmetric solutions \cite{gmr} but the construction does not tell us which of these solutions are bound states. Since these bound states are the microstates of the 3-charge extremal black hole, it is very important to be able to select the bound states out of all the possible supersymmetric solutions. The above conjecture says that those states are bound which do not have any `drift' type modes of excitations, and the others are unbound. If this conjecture is true, then we have in principle a way to identify  all 3-charge black hole states.

\bigskip

Note: \quad We will use the term `supertube' or just `tube' for 2-charge bound states in all duality frames, and at all values of the coupling. The supertube made from D0,NS1 charges carries a D2 dipole  charge, and we will call this the D0-NS1 supertube or the D2 supertube. 
When we use charges NS1,P we will call the object the NS1-P supertube.

\section{The NS1-P bound state in flat space}\label{then}
\setcounter{equation}{0}

We will find that the most  useful representation of the 2-charge system will be NS1-P.  We compactify a  circle $S^1$ with radius $R_y$; let $X^1\equiv y$ be the coordinate along this $S^1$. The elementary string (NS1) is wrapped on this $S^1$ with winding number $n_1$, and $n_p$ units of momentum run along the $S^1$. We are interested in the {\it bound} state of these charges. This corresponds to the NS1 being a single `multiwound' string with wrapping number $n_1$, and the momentum is carried on this NS1 by its transverse oscillations.\footnote{The momentum can also be carried by the fermionic superpartners of these oscillation degrees of freedom, but we will not focus on the fermions in what follows. For a discussion of  fermion modes in the 2-charge system see for example \cite{marika} .} 

\begin{figure}[htbp]
\vskip -1.0truecm
\begin{center}
\hskip - 1truein \includegraphics[width=4.5in,angle=270]{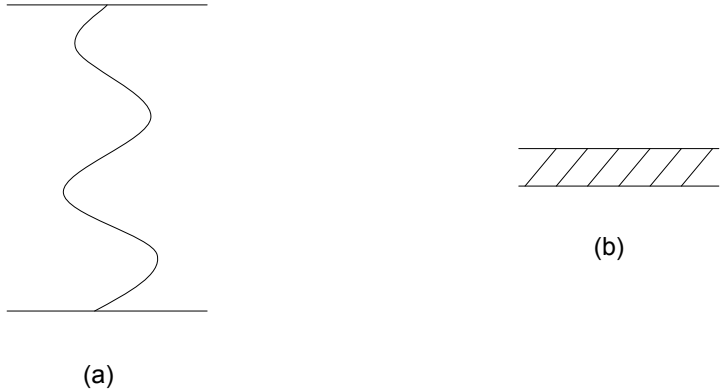}
\end{center}
\vskip -1.7truein
\caption[x] {{\small  (a) The NS1 carrying a transverse oscillation profile in the
covering space of $S^1$. (b) The strands of the NS1 as they
appear in the actual space.}}
\label{fig1m}
\end{figure}

Consider first the BPS states of this system. Then all the excitations carry momentum in one direction; we set this to be  the positive $y$ direction and  call these excitations `right moving'. In 
Fig.\ref{fig1m}(a) we  open up the multiwound string to its covering space where we can see the transverse oscillation profile. As explained in \cite{lm2} these  oscillations  cause the $n_1$ strands to separate from each other and the bound state acquires a transverse `size'. For the generic state of this 2-charge system the radius of the state is $\sim\sqrt{\alpha'}$ and the surface area of this region gives the entropy of the state by a Bekenstein type relation \cite{lm5}
\be
{A\over 4G}\sim S_{micro}=2\pi\sqrt{2}\sqrt{n_1n_5}
\ee
To understand the generic state better it is useful to look at configurations that have a much larger transverse size, and later take
the limit where we approach the generic state.  The relevant limit is explained in \cite{fuzz}. In this limit the wavelength of the vibration on the multiwound string is much larger than the radius of the $S^1$, so locally the strands of the NS1 look like Fig.\ref{fig1m}(b).  In the classical limit $n_1n_p\r\infty$ these strands will form  a continuous `strip', which will be described by (i) the profile of the strip in the space transverse to the $S^1$ and (ii) the `slope' of the strands at any point along the profile.\footnote{Note that the separation between successive strands is determined by the slope, since the radius of the $S^1$ is fixed at $R_y$.}

An S-duality gives NS1-P $\r$ D1-P, and a further T-duality along $y$ gives D1-P $\r$ D0-NS1. But note that locally the string is slanted, and the T-duality also generates a local $D2$ charge. Thus we get a `supertube'\footnote{In \cite{supercurve} the same dualities were performed in the reverse order.} where the D0-NS1 have formed a D2 \cite{supertube}. There is of course no {\it net} D2 charge; rather the D2 is a `dipole' charge. Note also that the slope of the NS1 in the starting NS1-P configuration implies that the momentum is partly along the direction of the `strip'. Since we do no dualities in the strip direction we will end up with momentum being carried along the D2 supertube. 

In this BPS configuration the D2 supertube is stationary. If we add some extra energy to the tube (while keeping its true (i.e. non-dipole) charges fixed) then we will get the dynamics of the supertube. But we can study the dynamics  in the NS1-P picture
and dualize to the D2 supertube at the end if we wish. In the NS1-P picture we just have to study a free, classical string.  Here  the left and right movers decouple and the problem can be solved exactly. Let us review this solution and extract the dynamics of the `supertube' in the limit of large charges.

\subsection{The classical string solution}

The string dynamics in flat space is described by the Nambu-Goto action
\be
S_{NG}=-T\sqrt{-\det[{\p X^\mu\over \p\chi^a}{\p X_\mu\over \p\chi^b}]}
\label{ng}
\ee
where
\be
T={1\over 2\pi\alpha'}
\ee
We can get an equivalent dynamics by introducing an auxiliary metric on the world sheet (this gives the Polyakov action)
\be
S_P=-{T\over 2}\int d^2 \chi\sqrt{-g}{\p X^\mu\over \p\chi^a}{\p X_\mu\over \p \chi^b} g^{ab}
\label{polyakov}
\ee
The variation of $g_{ab}$ gives
\be
{\p X^\mu\over \p\chi^a}{\p X_\mu\over \p \chi^b}-{1\over 2} g_{ab}{\p X^\mu\over \p\chi^c}{\p X_\mu\over \p \chi^d}g^{cd}=0
\ee
so $g_{ab}$ must be proportional to the induced metric. Substituting this $g_{ab}$ in (\ref{polyakov}) we get back (\ref{ng}), thus showing that the two actions are classically equivalent.

The $X^\mu$ equations give
\be
\p_a[\sqrt{-g}{\p X_\mu\over \p \chi^b} g^{ab}]=0
\ee
Note that the solution for the $X^\mu$ does not depend on the conformal factor of $g_{ab}$. 

We choose coordinates $\chi^0\equiv {\hat\tau},\, \chi^1\equiv {\hat\sigma}$ on the world sheet so that $g_{ab}=e^{2\rho}\eta_{ab}$ for some $\rho$. Writing
\be
\chi^+=\chi^0+\chi^1, ~~~\chi^-=\chi^0-\chi^1
\ee
we have
\be
g_{++}=0, ~~~g_{--}=0
\label{conformal}
\ee
Since the induced metric must be  proportional to $g_{ab}$ we get 
\be
{\p X^\mu\over \p\chi^+}{\p X_\mu\over \p \chi^+}=0, ~~~{\p X^\mu\over \p\chi^-}{\p X_\mu\over \p \chi^-}=0
\label{condition}
\ee
Thus in these coordinates we get a solution if the $X_\mu$ are harmonic functions
\be
X_\mu^{,a}{}_{a}=0
\label{xsol}
\ee
and they  satisfy (\ref{condition}). The equations (\ref{xsol}) imply that the coordinates $X^\mu$ can be expanded as
\be
X^\mu=X^\mu_+(\chi^+)+X^\mu_-(\chi^-)
\ee

We can use the residual diffeomorphism symmetry to set the harmonic function $X^0$ to\footnote{Note that it is more conventional to set a light cone coordinate $X^+$ to be linear in ${\hat\tau}$. Using a light cone coordinate allows the constraints (\ref{condition}) to be solved without square roots, but for us this is not important since we will not need to quantize the string.} 
\be
X^0={\hat a}+{\hat b}{\hat\tau}={\hat a}+{\hat b}\h(\chi^++\chi^-)
\label{btau}
\ee
Let the coordinate along the $S^1$ be called $y$.  We can solve the constraints to express the terms involving $y$ in terms of the other variables. We find
\be
\partial_+ y_+=\pm \sqrt{{{\hat b}^2\over 4}-\partial_+ X_+^i\partial_+ X_+^i}\,,\quad \partial_- y_-=\pm \sqrt{{{\hat b}^2\over 4}-\partial_- X_-^i\partial_- X_-^i}
\label{ysolex}
\ee
where $X^i, i=1\dots 8$ are the spatial directions transverse to the $S^1$.
The parameter $\hat b$ should be chosen in such a way that the coordinate $y$ winds $n_w$ times around a circle
of length $R_y$ when ${\hat\sigma}\to {\hat\sigma}+2\pi$.  There is no winding around any other direction. We also use a reference frame in which the string has no momentum in any direction transverse to the $S^1$. We let $0\le {\hat\sigma} < 2\pi$. Then the target space coordinates can be expanded as
\bea
y&=&{\alpha' n_p\over R_y}{\hat\tau} +n_w\,R_y\,{\hat\sigma}+\sum_{n\not=0}(c_n\,e^{i \,n\,\chi^-}+ d_n e^{i \,n\,\chi^ +})\nn
X^i&=&\sum_{n\not=0}(c^i_n\,e^{i \,n\,\chi^-}+ d^i_n e^{i \,n\,\chi^ +})
\label{xsolp}
\eea
Define
\be
S_+=\sqrt{{{\hat b}^2\over 4}-\partial_+ X_+^i\partial_+ X_+^i}\,,\quad S_-=\sqrt{{{\hat b}^2\over 4}-\partial_- X_-^i\partial_- X_-^i}
\ee
From the energy and winding required of the configuration we find
 that the choice of signs in (\ref{ysolex}) should be
\be
\p_+y_+=S_+, ~~~~\p_-y_-=-S_-
\ee

After an interval 
\be
\Delta {\hat\tau}=\pi
\label{ptau}
\ee 
all the $X^i$ return to their original values. This can be seen  by noting that ${\hat\sigma}$ is only a parameter that labels world sheet points, so the actual configuration of the system does not depend on the origin we choose for ${\hat\sigma}$. Thus consider  the change
\be
({\hat\tau}={\hat\tau}_1, ~{\hat\sigma}) ~~\r ~~ ({\hat\tau}={\hat\tau}_1+\pi, ~{\hat\sigma}+\pi)
\label{shift}
\ee 
From (\ref{xsolp}) we see that the $X^i$ are periodic with period ${\hat\tau}=\pi$. The coordinate $y$ does not  return to its original value, but in the classical limit that we have taken to get the `supertube' we have smeared over this direction and 
so the {\it value} of $y$ is not involved in describing the configuration of the supertube. But the {\it slope} of the NS1 at a point in the supertube {\it is} relevant, and is given by
\be
s={|{\p X^i\over \p {\hat\sigma}}|\Big / ({\p y\over \p{\hat\sigma}})}
\label{slope}
\ee
But
\be
{\p y\over \p {\hat\sigma}}={n_w R_y} + \sum_{n\ne 0} ~[(-in)c_n e^{in({\hat\tau}-{\hat\sigma})}+ (in) 
{d}_{n} e^{in({\hat\tau}+{\hat\sigma})}]
\ee
We see that $\p y/\p {\hat\sigma}$ is periodic under (\ref{shift}) and thus so is (\ref{slope}).

From (\ref{btau}) we see that when ${\hat\tau}$ changes by the above period then
\be
\Delta X^0={\hat b}\Delta {\hat\tau}={\hat b}\pi
\ee
and the supertube configuration returns to itself. 
But
\be
{\hat b}=\alpha' P^0\equiv \alpha' E
\ee
where $E$ is the energy of the configuration. We therefore find that the motion of the supertube is periodic
in the target space time coordinate with period
\be
\Delta t=\Delta X^0=\alpha' \pi E
\label{period}
\ee
For the NS1-P system the dipole charge is NS1 -- this arises from the fact that the NS1 slants as shown in Fig.\ref{fig1m}(b) and so 
there is a local NS1 charge along the direction of the supertube. The tension of the NS1 is $T=1/(2\pi\alpha')$. This is thus the mass of the dipole charge per unit length
\be
m_d={1\over 2\pi\alpha'}
\ee
We then see that (\ref{period}) can be recast as 
\be
\Delta t={1\over 2} {E\over m_d}
\label{periodp}
\ee
This form for the period will be of use to us later, because we will find that it holds in other duality frames as well.

\subsection{The linearized perturbation}

We can solve the NS1-P system exactly and we have thus obtained the exact dynamics of the supertube in flat space. For some purposes it will be useful to look at the small perturbations to the stationary tube configurations. We now study these small perturbations, starting in a slightly different way from the above analysis.

Consider first the string in a BPS configuration: The wave on the string is purely right moving. We know that in this case the waveform travels with the speed of light in the positive $y$ direction. Let us check that this is a solution of our string equations. This time we know the solution in the {\it static} gauge on the worldsheet: 
\be
t= b \t\tau, ~~~y=b\t\sigma
\label{static}
\ee
Writing $\xi^\pm=\t\tau\pm\t\sigma$ and noting that a right moving wave is a function of $\xi^-$ we expect the following to be a solution
\be
t=b\,{\xi^+ + \xi^-\over 2}\,,\quad y=b\,{\xi^+ - \xi^-\over 2}\,,\quad
X^i=X^i(\xi^-)
\label{left}
\ee
In these worldsheet coordinates the induced metric is
\be
ds^2=-b^2\,d\xi^+\,d\xi^-+({X^i}'{X^i}')\,(d\xi^-)^2
\ee
so it does not satisfy (\ref{conformal}). (Here prime denotes differentiation with respect to $\xi^{-}$). However we can change to new coordinates on the worldsheet
\be
(\tilde{\xi}^+,\xi^-)=(\xi^+ - f(\xi^-), \xi^-)
\label{change1}
\ee  
with 
\be
 f'(\xi^-) ={({X^i}'{X^i}') (\xi^-)\over b^2}
\ee
This brings the metric to the conformally flat form
\be
ds^2=-b^2\,d\xi^-\,d\tilde{\xi}^+
\ee
Moreover, rewriting (\ref{left}) in terms of $(\tilde{\xi}^+,\xi^-)$
\be
t=b\,{{\tilde{\xi}^+ + \xi^- + f(\xi^-)\over 2}}\,,\quad y=b\,{{\tilde{\xi}^+ - \xi^- + f(\xi^-)\over 2}}\,,\quad
{{X^i}}={{X^i}}(\xi^-)
\ee
one sees that the configuration is of the form
\be
X^\mu=x_+^\mu(\t\xi^+)+x_-^\mu(\xi^-)
\label{sum}
\ee
so that the $X^\mu$ are harmonic in the coordinates  $(\tilde{\xi}^+,\xi^-)$. Thus the coordinates $(\tilde{\xi}^+,\xi^-)$
are  conformal coordinates for the problem  and we have verified that (\ref{left}) is a solution of the equations of motion.

We now proceed to adding a small right moving perturbation, which was our goal. Consider the perturbed configuration
\bea
&&t=b\,{\tilde \tau}=b\,{{\tilde{\xi}^+ + \xi^- + f(\xi^-)\over 2}}\,,\quad y=b\,{\tilde \sigma}=
b\,{{\tilde{\xi}^+ - \xi^- + f(\xi^-)\over 2}}\nonumber\\
&&{{X^i}}={{X^i}}(\xi^-)+{{x^i}}(\tilde{\xi}^+)
\label{solFP}
\eea
where $x^i$ is assumed small. Then the induced metric on the worldsheet is
\be
ds^2 = -(b^2-2 {X^i}'{x^i}')\,d\xi^-\,d\tilde{\xi}^+ + O({ x}')^2\,(d\tilde{\xi}^+)^2
\ee
so that it is conformally flat to first order in the perturbation. The target space coordinates $X^\mu$ in (\ref{solFP}) are
clearly of the form (\ref{sum}) so they are harmonic, and we have found a solution of the string equations of motion.

\subsection{Summary}

We  can get a general solution of the NS1-P system by taking arbitrary harmonic functions $X^i$ in (\ref{xsolp}) and determining $X^0$, $y$ by (\ref{btau}), (\ref{ysolex}). Taking the classical limit where the strands of the string forms a continuum gives the arbitrary motion of the supertube, and the period of this motion is given by (\ref{periodp}). The conformal gauge coordinate 
$\hat\sigma$ that is used on the string is not very intuitive, since it is determined by the state of the string. We next looked at the linearized perturbation to a BPS state, and this time we started with an intuitively simple coordinate on the string -- the static gauge coordinate $\t\sigma$ proportional to the spacetime coordinate $y$. We found the explicit map (\ref{change1}) to the conformal gauge coordinates. The solution to the linearized problem was then given by an arbitrary choice of the ${x^i}$ in (\ref{solFP}).

We will now see that these solutions reproduce the behavior of the D2 brane supertube at the exact and linearized levels respectively. The NS1-P system is the easiest way to solve the problem, since it exhibits the separation of the dynamics into a left and a right mover; this separation is not obvious in the other duality frames.

\section{Perturbations of the D0-NS1 supertube} \label{secpert}
\setcounter{equation}{0}

In this section we will consider the more conventional definition of the supertube: The true charges are D0-NS1 and the dipole charge is a D2. The dynamics is given by the DBI action of the D2 with worldvolume fields corresponding to the true charges.  In \cite{pm} perturbations were considered around the `round supertube' which has as its profile a circle in the $(X_1,X_2)$ plane. This supertube has the maximum possible angular momentum $J$ for its charges.  So if we add a small perturbation to it we know that we will  not get a  `drift' through a set of supertube configurations --  $J$ is conserved and   there are no other configurations with this value of $J$. So even though periodic excitations were found for this supertube we cannot conclude from this that small perturbations to the generic supertube will also be periodic. Thus we wish to extend the computation of 
\cite{pm} to the generic supertube. We will write the equations of motion for the generic case, but instead of solving them directly we will note that we have already solved the problem in NS1-P language and we will just dualize the solution there and check that it solves the equations for the D0-NS1 supertube.\footnote{The fact
that for given charges there is a range of possible configurations  around a generic supertube was also 
noted in \cite{bho}.}

We work in flat space with a compact $S^1$ of length ${\tilde L}_y=2\pi {\tilde R}_y$, 
parametrized by the coordinate $y$.  We have already obtained the general motion of the supertube in the NS1-P description, and
below will verify that this solves the general D2 supertube equations as well.  But first we check the behavior of small perturbations, and for this purpose we model our presentation as close to that of \cite{pm} as possible. Thus we let the supertube lie along a closed curve $\gamma$ in the $(X_1,X_2)$ plane, but $\gamma$ need not be a circle as in \cite{pm}. The worldvolume of the D2 will be  $\gamma\times S^1$.

Let $R$ and $\sigma$ be the radial and angular coordinates in the $(X_1,X_2)$   plane. We will
denote by $Z_a$ all the coordinates other than $t, X_1, X_2, y$. We will also sometimes use the notation
${\bf X}^I=\{X_1,X_2,Z_a\}$. We fix a gauge in which the world volume coordinates on the D2 brane are 
$t,\sigma, y$. Thus the angular coordinate in the supertube plane serves as the parameter along the supertube curve $\gamma$.  On the D2 world volume we have a gauge field, for which we adopt the gauge
\be
A_t=0
\ee
Thus the gauge field has the form
\be
A=A_\sigma\,d\sigma+A_y\,dy
\ee

The D2-brane Lagrangian density is given by usual Born-Infeld term:
\be
\mathcal{L}=-T_2\,\sqrt{-\mathrm{det}(g+F)}
\label{bi1}
\ee
where $T_2$ is the D2 brane tension, $g$ is the metric induced on the D2 world volume and $F$ is the field 
strength of $A$. There are no  background fields, so there is no Chern-Simons term in the action.

We want to consider fluctuations around a static configuration described by the curve
\be
R={\bar R}(\sigma)\,,\quad Z_a=0
\ee
and field strength
\be
F={\bar E}\,dt\wedge dy + {\bar B}(\sigma)\,dy \wedge d\sigma
\ee

It is known \cite{supertube} that this configuration\footnote{Since the  configuration  is independent of $t,y$, the Bianchi identity requires that $\bar E$ be a constant. There is no restriction on $\bar B$ and it is an arbitrary function of $\sigma$.} satisfies the equations of motion and is supersymmetric for arbitrary 
${\bar R}(\sigma),{\bar B}(\sigma)$ if ${\bar E}^2=1$ and  $\mathrm{sign}\, {\bar B}(\sigma)=\pm 1$.
Without any loss of generality, we will take ${\bar E}=1$ and  $\mathrm{sign}\, {\bar B}(\sigma)=1$ in what follows. The electric field 
$\bar E$ induces a NS1 integer charge given by
\be
n_1={1\over T}\,\int d\sigma\, \Pi_y = 
{1\over T}\,\int d\sigma\, {\partial \mathcal{L}\over \partial (\partial_t A_y)}
\ee 
where $T$ is the NS1 tension. The magnetic field $\bar B$ induces a D0 integer charge equal to
\be
n_0={T_2\over T_0}\,\int dy d\sigma\,{\bar B}(\sigma)
\ee
where $T_0$ is the D0 brane mass. 

We want to study fluctuations around the configuration described above. So we expand the Lagrangian up to quadratic
order. We will assume that the fluctuations do not depend on $y$. We parametrize the D2-brane world volume 
as
\be
R={\bar R}(\sigma)+r(\sigma,t)\,,\quad Z_a=z_a(\sigma,t)
\ee
and the field strength as
\bea
F&=&E\,dt\wedge dy + B\,dy\wedge d\sigma+\partial_t a_\sigma\,dt \wedge d\sigma\nonumber\\
&=&({\bar E}+\partial_t a_y)\,dt\wedge dy + ({\bar B}(\sigma)-\partial_\sigma a_y)\,dy \wedge d\sigma + \partial_t a_\sigma\,dt\wedge d\sigma
\eea
where lower case quantities denote the fluctuations. The metric induced on the D2 brane world volume is 
\be
ds^2=-dt^2 + (\partial_\sigma {\bar R}\,d\sigma + \partial_\sigma r\, d\sigma + \partial_t r\,dt)^2 +
({\bar R}(\sigma)+r)^2\,d\sigma^2 +dy^2 + (\partial_\sigma z_a\,d\sigma + \partial_t z_a\,dt)^2
\ee
The Lagrangian density $\mathcal{L}$ for the system is given by
\bea
\!\!-\frac{\mathcal{L}}{T_2}&\!\!\!=\!\!\!&\sqrt{-{\rm det}(g+F)} \nonumber \\
\!\!&\!\! = \!\!\!&\{\, -|\partial_t {\bf X}|^2 |\partial_\sigma {\bf X}|^2 + 
(\partial_t {\bf X}\cdot \partial_\sigma {\bf X})^2+(1-E^2)|\partial_\sigma {\bf X}|^2 + B^2(1-|\partial_t {\bf X}|^2)\nonumber\\
\!\!&&\quad-2 E B (\partial_t {\bf X}\cdot \partial_\sigma {\bf X})-(\partial_t a_\sigma)^2\, \}^{1/2}\nonumber\\
\!\!&\!\!\!=\!\!\!&\{-R^2 [(\partial_t R)^2+|\partial_t z_a|^2]-
(\partial_t R)^2 |\partial_\sigma z_a|^2 -
(\partial_\sigma R)^2 |\partial_t z_a|^2+2 \partial_t R \partial_\sigma R \,\partial_t z_a \partial_\sigma z_a \nonumber\\
\!\!&&\quad +(1-E^2)
[R^2+(\partial_\sigma R)^2+|\partial_\sigma z_a|^2]+B^2 [1-(\partial_t R)^2-|\partial_t z_a|^2]\nonumber\\
\!\!&&\quad- 2 E B [\partial_t R \partial_\sigma R+\partial_t z_a\,\partial_\sigma z_a] +
(\partial_t z_a\,\partial_\sigma z_a)^{2}-|\partial_\sigma z_a|^2|\partial_t z_a|^2 \nonumber\\
\!\!&&\quad -(\partial_t a_\sigma)^2\}^{1/2}
\label{lag}
\eea

We wish to find the equations of motion up to linear order in the perturbation. To do this we expand $\mathcal{L}$ up 
to second order in $r,a_y,a_\sigma$:
\be
\frac{\mathcal{L}}{T_2}=L^{(0)}+L^{(1)}+L^{(2)}
\ee
We find
\be
L^{(0)}= -{\bar B}
\ee
\be
L^{(1)}=\Bigl[\partial_\sigma a_y + {{\bar R}^2+(\partial_\sigma {\bar R})^2\over {\bar B}}\,
\partial_t a_y +\partial_\sigma {\bar R}\,\partial_t r \Bigr]
\ee

We see that at first order in the perturbation the  Lagrangian reduces to a total derivative in 
$\sigma$ and $t$; this verifies the fact that our  starting configuration satisfies the equations of motion. The term quadratic in the perturbation is
\bea
L^{(2)}&\!\!=\!\!&-{1\over 2} \,\Bigl[ -{{\bar R}^2+(\partial_\sigma {\bar R})^2+{\bar B}^2\over {\bar B}}\,(\partial_t r)^2 -
2 \partial_t r\,\partial_\sigma r \nonumber\\
\!\!&-\!\!&2\,{{\bar R}^2+(\partial_\sigma {\bar R})^2+{\bar B}^2\over {\bar B}^2}\,\partial_\sigma {\bar R}\,\partial_t r\,\partial_t a_y
-4 {\partial_\sigma {\bar R}\over {\bar B}}\,\partial_\sigma r\,\partial_t a_y-4 {{\bar R}\over {\bar B}}\,r\,\partial_t a_y\nonumber\\
\!\!&-\!\!&{({\bar R}^2+(\partial_\sigma {\bar R})^2+{\bar B}^2)\,({\bar R}^2+(\partial_\sigma {\bar R})^2)\over {\bar B}^3}\,(\partial_t a_y)^2 -2 
{{\bar R}^2+(\partial_\sigma {\bar R})^2\over {\bar B}^2}\,\partial_t a_y\,\partial_\sigma a_y \nonumber\\
\!\!&-&\!\!{{\bar R}^2+(\partial_\sigma {\bar R})^2+{\bar B}^2\over {\bar B}}\,(\partial_t z_a)^2 -
2 \partial_t z_a\,\partial_\sigma z_a 
-{|\partial_t a_\sigma|^2\over {\bar B}}
\Bigr]
\eea

From this Lagrangian we find the following equations of motion for the linearized perturbation:
\bea
&&\!\!\!\!\!\!\!\!\!\!
{{\bar R}^2+(\partial_\sigma {\bar R})^2+{\bar B}^2\over {\bar B}}\,\partial_t^2 r + 2\partial_t\partial_\sigma r +
\Bigl({{\bar R}^2+(\partial_\sigma {\bar R})^2+{\bar B}^2\over {\bar B}}\,\partial_t^2 a_y + 2\partial_t\partial_\sigma a_y\Bigr)\,
{\partial_\sigma {\bar R}\over {\bar B}}\nonumber\\
&&\qquad\qquad\qquad\qquad\qquad\qquad\qquad\qquad\qquad\qquad\qquad\quad\,\,-2{{\bar R}\over {\bar B}}\,\partial_t a_y + 2\partial_\sigma \Bigl({\partial_\sigma {\bar R}\over {\bar B}}\Bigr)\,\partial_t a_y~=~0
\nonumber\\
&&\!\!\!\!\!\!\!\!\!\!\Bigl({{\bar R}^2+(\partial_\sigma {\bar R})^2+{\bar B}^2\over {\bar B}}\,\partial_t^2 a_y + 2\partial_t\partial_\sigma a_y\Bigr)\,
{{\bar R}^2+(\partial_\sigma {\bar R})^2\over {\bar B}^2}\nonumber\\ 
&&~~~~~~~~+\Bigl({{\bar R}^2+(\partial_\sigma {\bar R})^2+{\bar B}^2\over {\bar B}}\,\partial_t^2 r + 2\partial_t\partial_\sigma r\Bigr)\,
{\partial_\sigma {\bar R}\over {\bar B}}+2{{\bar R}\over {\bar B}}\,\partial_t r +
\partial_\sigma\Bigl( {{\bar R}^2+(\partial_\sigma {\bar R})^2\over {\bar B}^2}\Bigr)\,\partial_t a_y~=~0\nonumber\\
&&\!\!\!\!\!\!\!\!\!\!{{\bar R}^2+(\partial_\sigma {\bar R})^2+{\bar B}^2\over {\bar B}}\,\partial_t^2 z_a + 2\partial_t\partial_\sigma z_a~=~0\nonumber\\
&&\!\!\!\!\!\!\!\!\!\!\partial_t^2 a_\sigma~=~0
\label{eom}
\eea
We have an additional equation coming from the variation of $A_t$; this is the Gauss law which says
\be
\p_\sigma E_\sigma\equiv \partial_\sigma\partial_t a_\sigma =0
\label{gauss}
\ee
The last equation in (\ref{eom}) and (\ref{gauss}) together say that we can add an electric field along the $\sigma$ direction but this field will be constant in both $\sigma$ and $t$. We will henceforth set this additional $E$ to zero, and thus
$a_\sigma=0$ for the rest of the calculation.

 Note that only time derivatives of fields occur in the equations; there are no terms where the fields appear without such time derivatives.   Thus any time independent perturbation is a solution to the equations. This tells us that we can make arbitrary time independent deformations of the supertube, reproducing the known fact that the supertube has a family of time independent solutions.

The D0 and NS1 integer charges of the perturbed configuration are
\bea
n_0&=&{T_2\over T_0}\,\int dy d\sigma\,({\bar B}(\sigma)-\partial_\sigma a_y)=
{T_2\over T_0}\,\int dy d\sigma\,{\bar B}(\sigma)\nn
n_1&=&{T_2\over T}\,\int d\sigma\,\Bigl[{{\bar R}^2+(\partial_\sigma {\bar R})^2\over {\bar B}}+ 2{{\bar R}\over {\bar B}}\,r\nonumber\\
~~~~&&+{({\bar R}^2+(\partial_\sigma {\bar R})^2)({\bar R}^2+(\partial_\sigma {\bar R})^2+{\bar B}^2) \over {\bar B}^3}\,\partial_t a_y+
{{\bar R}^2+(\partial_\sigma {\bar R})^2\over {\bar B}^2}\,\partial_\sigma a_y\nonumber\\
~~~~&&+{{\bar R}^2+(\partial_\sigma {\bar R})^2+{\bar B}^2\over {\bar B}^2}\,\partial_\sigma {\bar R}\,\partial_t r+2 {\partial_\sigma {\bar R}\over {\bar B}}\,
\partial_\sigma r
\Bigr]
\eea

We see that the D0 charge is unchanged by the perturbation. This charge in fact is a topological invariant of the gauge field configuration. For the NS1 charge we can check conservation by explicitly computing the time derivative and verifying that it vanishes.

The angular momentum in the $(X_1,X_2)$ plane is
\be
J= \int d\sigma\,dy\,(\Pi_2 X_1- \Pi_1 X_2)
\ee
where
\be
\Pi_i = {\partial \mathcal{L}\over \partial (\partial_t X_i)}\,\,,\,\,i=1,2
\ee
From the Lagrangian (\ref{lag}) we find
\be
\Pi_i=-T_{2}^{2}\left({\partial_t X_i \,[(\partial_\sigma {\bf X})^2+B^2]-\partial_\sigma X_i\, 
[(\partial_t {\bf X} \partial_\sigma {\bf X})-E B]\over \mathcal{L}}\right)
\ee
Expanding $J$ up to first order in the perturbation we get
\be
J=T_2\,(2\pi {\tilde R}_y)\,\int d\sigma\,\Bigl[{\bar R}^2+ 2 {\bar R} \,r +{{\bar R}^2\,({\bar R}^2+(\partial_\sigma {\bar R})^2+{\bar B}^2)\over {\bar B}^2}\,\partial_t a_y\Bigr]
\ee
 
\subsection{Using the NS1-P solution: Linear perturbation }

The equations (\ref{eom}) for the perturbations to the D2 brane look complicated, but we will obtain the solution by dualizing the NS1-P solution found above. Recall that we have split the spatial coordinates transverse to the $S^1$ (i.e. the ${\bf X}$)   as ${\bf X}=\{X_1,X_2,Z_a\}$. To arrive at the D2 supertube in the $(X_1,X_2)$ plane we assume  that the right moving wave on the NS1 has its transverse oscillations only in the $(X_1, X_2)$ plane. This solution is then perturbed by a small
left-moving wave. Recall that we had defined static gauge coordinates $\t\tau, \t\sigma$ (\ref{static}) on the world sheet and then obtained the conformal coordinates $\t\xi^+, \xi^-$. We will find it convenient to use as independent variables $\t\tau$ and $\xi^-$. This is because from (\ref{static}) we see that $\t\tau$ directly gives the target space time $t$, and $\xi^-$ is the variable in terms of which we have the basic right moving wave ${\bf X(\xi^-)}$ that gives the unperturbed solution. Thus we have
\be
\tilde{\xi}^+=\xi^+-f(\xi^-)=2\tilde\tau-\xi^- -f(\xi^-)\,,\quad ~~~ f'={(X_1')^2+(X_2')^2\over b^2}
\ee
For the NS1-P solution the target space coordinates are given by
\bea
t&=&b\,{\tilde \tau}\nn
y&=&b\,{\tilde \sigma}=b\,(\tilde\tau-\xi^-)\nonumber\\
X_i(\xi^-,\tilde \tau)&=&X_i(\xi^-)+x_i(\tilde{\xi}^+)\,\,,\,\,i=1,2\nn
Z_a(\xi^-,\tilde\tau)&=&z_a(\tilde{\xi}^+)
\label{pertsol}
\eea
where $x_i, z_a$ are small perturbations.

We perform an S-duality to go from NS1-P to D1-P, and then a T-duality along $S^1$ to get the D0-NS1 supertube. The $S^1$ coordinate $y$ goes, under these changes, to the component $A_y$ of the gauge field on the D2. In the normalization of the gauge field $A$ used in the action (\ref{bi1}) we just get
\be
y~\r~A_y
\ee
so from (\ref{pertsol}) we have
\be
A_y=t-b\,\xi^-
\ee

In this solution derived by duality from NS1-P the natural coordinates on the D2 are $(\t\tau, \xi^-, y)$.\footnote{In these coordinates we can see that the electric field is $E=\partial_t A_y=1$, as expected for the stationary supertube configurations.} But when we wrote the DBI action for the D2 the natural coordinates were $(t, \sigma, y)$, where $\sigma$ was the angle in the $(X_1,X_2)$ plane
\be
\tan \sigma = {X_2(\xi^-,\tilde\tau)\over X_1(\xi^-,\tilde\tau)}
\label{sigma-xi}
\ee
The coordinates $t$ and $\tilde \tau$ are related by a constant, so there is no difficulty in replacing the $t$ by $\t\tau$ in converting the NS1-P solution to a D0-NS1 supertube solution. But the change $\xi^-\r\sigma$ is more complicated, and will necessitate the algebra steps below. 
Inverting (\ref{sigma-xi}) gives
\be
\xi^-=\xi^-(\sigma,\tilde\tau)
\label{reverse}
\ee
so we see that the change $\xi^-\r\sigma$ depends on time as well, if the supertube is oscillating. The variables describing the  supertube configuration will be 
\bea
R(\sigma,\tilde\tau)&=&\sqrt{X_1^2(\xi^-(\sigma,\tilde\tau),\tilde\tau)+X_2^2(\xi^-(\sigma,\tilde\tau),\tilde\tau)}\nonumber\\
A_y(\sigma,\tilde\tau)&=&t-b\,\xi^-(\sigma,\tilde\tau)\,\nn
Z_a(\sigma,\tilde\tau) &=& Z_a(\xi^-(\sigma,\tilde\tau),\tilde\tau) 
\label{fullsol}
\eea
which should satisfy the equations for the D0-NS1 supertube.

First consider the unperturbed configuration. Here the transformation (\ref{reverse}) does not depend on $\t\tau$. For the variables of the unperturbed configuration we write
\be
\bar{X}_i=X_i(\bar{\xi}^-(\sigma))\,,\quad \bar{X}'_i = X'_i(\bar{\xi}^-(\sigma))\,\,,\,\,i=1,2
\ee
where the prime denotes a derivative with respect to the argument $\bar{\xi}^-$.
The function $\bar{\xi}^-(\sigma)$ will be the solution of the equation
\be
\tan\sigma={\bar X_2(\bar{\xi}^-)\over \bar X_1(\bar{\xi}^-)}
\ee
and the stationary configuration will be  given by
\bea
{\bar R}(\sigma)&=&\sqrt{\bar{X}_1^2+\bar{X}_2^2}\,\nn
{\bar B}(\sigma)&=&-\partial_\sigma A_y=b\,\partial_\sigma \bar{\xi}^-(\sigma)
\eea

From the above definitions we can derive the identities
\bea
{{\bar B}\over {\bar R}^2}&=&{b\over \bar{X}_1 \bar{X}'_2 - \bar{X}_2 \bar{X}'_1}\,\nn
\partial_\sigma {\bar R}&=& {{\bar B}\over b\, {\bar R}}\,(\bar{X}_1 \bar{X}'_1 + \bar{X}_2 \bar{X}'_2)
\label{identities}
\eea
Using these identities one can prove a relation that will be important in the following
\be
{\bar R}^2+(\partial_\sigma {\bar R})^2={\bar B}^2\,\bar{f}'
\label{id2}
\ee
where $\bar{f}'=f'(\bar{\xi}^-(\sigma))$.
Now consider the small perturbation on the supertube. We will keep all quantities to linear order in the  $x_i, z_a$.
Inverting the relation (\ref{sigma-xi}) gives us
\be
\xi^-=\bar{\xi}^-+\hat{\xi}^-\,,\quad 
\hat{\xi}^-(\sigma,\tilde\tau)=-{\bar{X}_1\,{\tilde x}_2 -\bar{X}_2\,{\tilde x}_1\over 
\bar{X}_1 \bar{X}'_2 - \bar{X}_2 \bar{X}'_1 }
\label{relation1}
\ee
where
\be
{\tilde x}_i=x_i(2\tilde\tau-\bar{\xi}^--f(\bar{\xi}^-))\,\,,\,\,i=1,2
\ee
Using (\ref{relation1}), the first identity  in (\ref{identities}), and performing an expansion to
first order in $x_i$, we find the 
perturbation around the static configuration 
\bea
r(\sigma,\tilde\tau)&=&R(\sigma,\tilde\tau)-{\bar R}(\sigma)={{\bar B}\over b\,{\bar R}}\,(\tilde{x}_1 \bar{X}'_2 - \tilde{x}_2 \bar{X}'_1)\nonumber\\
&=&\tilde{x}_1\,\cos\sigma+\tilde{x}_2\,\sin\sigma+{\partial_\sigma {\bar R}\over {\bar R}}\,
(\tilde{x}_1\,\sin\sigma-\tilde{x}_2\,\cos\sigma)\nonumber\\
a_y(\sigma,\tilde\tau)&=&-b\,\hat{\xi}^-=-{{\bar B}\over {\bar R}^2}\,(\tilde{x}_1 \bar{X}_2 - \tilde{x}_2 \bar{X}_1)
\nonumber\\
&=&-{{\bar B}\over {\bar R}}\,(\tilde{x}_1\,\sin\sigma-\tilde{x}_2\,\cos\sigma)\nonumber\\
z_a(\sigma,\tilde\tau)&=&z_a(2\tilde\tau-\bar{\xi}^--f(\bar{\xi}^-))~\equiv~ {\tilde z}_a
\label{pert}
\eea

We would like to check that the functions $r$, $a_y$ and $z_a$ defined above satisfy the equations of motion 
(\ref{eom}). For this purpose, some useful identities are

\be
\partial_\sigma {\tilde x}'_i=-{{\bar B}\over b}(1+\bar{f}')\,{\tilde x}''_i\,,\quad 
\partial_\sigma {\tilde z}'_a=-{{\bar B}\over b}(1+\bar{f}')\,{\tilde z}''_a
\ee
We can simplify some expressions appearing in the equations of motion (\ref{eom})
\bea
{{\bar R}^2+(\partial_\sigma {\bar R})^2+{\bar B}^2\over {\bar B}}\,\partial_t^2 r + 2\partial_t\partial_\sigma r
 &=& {4\over b}\,\Bigl[-\tilde{x}'_1\,\sin\sigma+\tilde{x}'_2\,\cos\sigma+{\partial_\sigma {\bar R}\over {\bar R}}\,
(\tilde{x}'_1\,\cos\sigma+\tilde{x}'_2\,\sin\sigma)\Bigr]\nonumber\\
&&\qquad\qquad+ {4\over b}\partial_\sigma 
\Bigl({\partial_\sigma {\bar R}\over {\bar R}}\Bigr)\,(\tilde{x}'_1\,\sin\sigma-\tilde{x}'_2\,\cos\sigma)\nonumber\\
{{\bar R}^2+(\partial_\sigma {\bar R})^2+{\bar B}^2\over {\bar B}}\,\partial_t^2 a_y + 2\partial_t\partial_\sigma a_y
&=&-4{{\bar B}\over b\,{\bar R}}\,(\tilde{x}'_1\,\cos\sigma+\tilde{x}'_2\,\sin\sigma)
\nonumber\\
&&\qquad -{4\over b}\partial_\sigma 
\Bigl({{\bar B}\over {\bar R}}\Bigr)\,(\tilde{x}'_1\,\sin\sigma-\tilde{x}'_2\,\cos\sigma)\nonumber\\
{{\bar R}^2+(\partial_\sigma {\bar R})^2+{\bar B}^2\over {\bar B}}\,\partial_t^2 z_a + 2\partial_t\partial_\sigma z_a&=&0
\eea

The last identity proves that the equations for $z_a$ are satisfied. For the equations involving 
 $r$ and $a_y$ some more work is needed.  
The l.h.s. of the first equation in (\ref{eom}) is equal to
\bea
&& {4\over b}(\tilde{x}'_1\,\sin\sigma-\tilde{x}'_2\,\cos\sigma)\Bigl[-1+ \partial_\sigma 
\Bigl({\partial_\sigma {\bar R}\over {\bar R}}\Bigr)-\partial_\sigma 
\Bigl({{\bar B}\over {\bar R}}\Bigr)\,{\partial_\sigma {\bar R}\over {\bar B}} +{{\bar B}\over {\bar R}}\Bigl( {{\bar R}\over {\bar B}} -
\partial_\sigma \Bigl({\partial_\sigma {\bar R}\over {\bar B}}\Bigr)\Bigr)\Bigr]\nonumber\\
&&+{4\over b} (\tilde{x}'_1\,\cos\sigma+\tilde{x}'_2\,\sin\sigma)\,\Bigl[{\partial_\sigma {\bar R}\over {\bar R}}-
{{\bar B}\over {\bar R}}\,{\partial_\sigma {\bar R}\over {\bar R}}\Bigr]
\eea
which, after some algebra, is seen to vanish. The l.h.s. of the second equation in (\ref{eom})
is
\bea
&& {4\over b}(\tilde{x}'_1\,\sin\sigma-\tilde{x}'_2\,\cos\sigma)\Bigl[-{{\bar R}^2+(\partial_\sigma {\bar R})^2\over {\bar B}^2}\,
\partial_\sigma \Bigl({{\bar B}\over {\bar R}}\Bigr)-{\partial_\sigma {\bar R}\over {\bar B}}
\Bigl(1-\partial_\sigma\Bigl({\partial_\sigma {\bar R}\over {\bar R}}\Bigr)\Bigr)\nonumber\\
&&\qquad +{{\bar R}\over {\bar B}}{\partial_\sigma {\bar R}\over {\bar R}}-{1\over 2}{{\bar B}\over {\bar R}}\partial_\sigma \Bigl({
{\bar R}^2+(\partial_\sigma {\bar R})^2\over {\bar B}^2}\Bigr)\Bigr]\nonumber\\
&&+{4\over b} (\tilde{x}'_1\,\cos\sigma+\tilde{x}'_2\,\sin\sigma)\,\Bigl[-{{\bar B}\over {\bar R}}{{\bar R}^2+(\partial_\sigma {\bar R})^2\over {\bar B}^2}+{(\partial_\sigma {\bar R})^2\over {\bar B} {\bar R}}+{{\bar R}\over {\bar B}}\Bigr]
\eea
which also vanishes.

We thus find that the expressions (\ref{pert}), with $x_i, z_a$ arbitrary functions of their arguments, satisfy the equations (\ref{eom}).

\subsection{Period of oscillation}

We would  like to determine the 
period of the oscillations of the solution (\ref{pert}).
The world sheet coordinate  $\tilde\sigma$  has a period $2\pi$. 
The time dependence of the solution (\ref{pert}) is contained in functions 
$x_i(2\tilde\tau-{\bar \xi}^--f({\bar \xi}^-))$ and $z_a(2\tilde\tau-{\bar \xi}^--f({\bar \xi}^-))$. The quantity $({\bf X}')^2({\bar \xi}^-)$ which appears
in the definition of $f({\bar \xi}^-)$ will be the sum of a constant term, ${\tilde R}^2$, plus terms periodic in ${\bar \xi}^-$:
\be
({\bf X}')^2({\bar \xi}^-)={\tilde R}^2+\sum_{n\not=0} (a_n e^{i n \,{\bar \xi}^-}+c.c.)\,
\ee
which implies that $f$ has the form
\be
f({\bar \xi}^-)={{\tilde R}^2\over b^2}\,{\bar \xi}^-+ \sum_{n\not=0} (b_n e^{i n\, {\bar \xi}^-}+c.c.)
\label{tilder}
\ee
The functions $x_i, z_a$ are functions of the coordinate ${\bar \xi}^-$ along the supertube. This 
supertube is a closed loop, so all functions on it are periodic under the shift $(\t\tau,{\bar \xi}^-)\r(\t\tau,{\bar \xi}^-+2\pi)$.  This implies
\be
x_i \Bigl(2\tilde\tau-{\bar \xi}^--f({\bar \xi}^-)\Bigr)= x_i \Bigl(2\tilde\tau-{\bar \xi}^--f({\bar \xi}^-) -2\pi \Bigl(1+ \frac{\t R^{2}}{b^{2}}\Bigr)\Bigr)
\ee  
where we have used (\ref{tilder}) to get the change in $f({\bar \xi}^-)$.

We have a similar relation for $z_a(2\tilde\tau-{\bar \xi}^--f({\bar \xi}^-))$. Thus the period of the oscillations is given by
\be
\Delta t=b\Delta \tilde{\tau}= \pi\,{b^2+{\tilde R}^2\over b}
\ee
This form of the period is similar to that found in \cite{pm}; it reduces to the period found there when the radius $\bar R$ is a constant.

To arrive at our more general form (\ref{periodp}) we write 
\be
{\bar \xi}^-+f({\bar \xi}^-)=\int_0^{{\bar \xi}^-} d\chi (1+f'(\chi))
\ee
So the change in ${\bar \xi}^-+f({\bar \xi}^-)$ when ${\bar \xi}^-$ increases by $2\pi$ can be written as
$\int_0^{2\pi}d\chi (1+f'(\chi))$. We then find that the argument of $x_i, z_a$ are unchanged when $(\t\tau, {\bar \xi}^-)\r (\t\tau+\Delta \t\tau, {\bar \xi}^-+2\pi)$ with
\be
2\Delta \tilde{\tau} -\int_{0}^{2\pi}
\left(1+ f'(\bar{\xi}^-)\right)d\bar{\xi}^- =0
\ee
Using the identity (\ref{id2}) we write the above as
\be
\Delta \tilde{\tau} = \frac{1}{2}\int_{0}^{2\pi}\left(1+ \frac{{\bar R}^2+(\partial_\sigma {\bar R})^2}{{\bar B}^{2}}
\right)d\bar{\xi}^- 
\ee

Now using the fact that ${\bar B}=b\,\partial_\sigma {\bar\xi}^-$, $\Delta t= b \Delta \tilde{\tau}$ and changing variables from
$\bar{\xi}^-$ to $\sigma$ we get
\be
\Delta t = \frac{1}{2}\int d\sigma \left( {\bar B} + \frac{{\bar R}^2 +(\partial_\sigma {\bar R})^2}{{\bar B}} \right)
\label{dt}
\ee

Now we express (\ref{dt}) in terms of the NS1 and D0 charges 
(we can use the unperturbed values of these quantities), using
\be
n_1={T_2\over T}\,\int\! d\sigma\,{{\bar R}^2+(\partial_\sigma {\bar R})^2\over {\bar B}} \ \ , \ \ n_0={T_2\over T_0}\,\int\! dy \,d\sigma\,{\bar B}
\ee

We get
\be
\Delta t = \frac{1}{2}\left(\frac{n_0 T_0 + n_1 T {\tilde L}_y}{T_2 {\tilde L}_y}\right)
\ee
where ${\tilde L}_y=2\pi {\tilde R}_y$ is the length of $y$ circle in the D0-NS1 duality 
frame.

Note that $n_0 T_0 + n_1 T {\tilde L}_y$ is the mass of the BPS state and since we have added only an infinitesimal perturbation it is to leading order the energy $E$ of the configuration. Further $T_2{\tilde L}_y$ is the mass of the D2 dipole charge per unit length of the supertube curve $\gamma$. Thus we see that the period again has the form (\ref{periodp})
\be
\Delta t={1\over 2} {E\over m_d}
\label{periodpp}
\ee

\subsection{Using the NS1-P solution: Exact dynamics }

Now consider the exact NS1-P solution (i.e. not perturbative around a BPS configuration). We again perform the required dualities to transform this solution into a solution of the D2 supertube. We will use as world-volume coordinates 
for the D2 brane $(\hat\tau,\hat\sigma, y)$.  Then the D2  solution is given by
\bea
&&X^i=X^i_+(\chi^+)+X^i_-(\chi^-)\,,\quad A_y=y_+(\chi^+)+y_-(\chi^-)\nonumber\\
&&E=\partial_{\hat\tau} A_y = \partial_+y_+ + \partial_- y_-=S_+-S_-\nonumber\\
&&B=-\partial_{\hat\sigma} A_y = -\partial_+y_+ + \partial_- y_-=-(S_++S_-)
\label{fullsolexact}
\eea
In this subsection $X^i$ denotes all coordinates other than $t$ and $y$.
We wish to prove that (\ref{fullsolexact}) satisfies the dynamical equations of the D2 brane. The DBI lagrangian density is given by
\bea
\frac{\mathcal{L}}{T_2}&=&-\sqrt{-{\rm det}(g+F)}\nonumber\\
&=&-[-(\partial_{\hat\tau} X)^2 (\partial_{\hat\sigma} X)^2 + 
(\partial_{\hat\tau} X \partial_{\hat\sigma} X)^2+(b^2-E^2)(\partial_{\hat\sigma} X)^2 + B^2(b^2-(\partial_{\hat\tau} X)^2)\nonumber\\&&\quad -2 E B (\partial_{\hat\tau} X \partial_{\hat\sigma} X)]^{1/2}
\eea
The equations of motion are
\bea
&&\partial_{\hat\tau}\Bigl[{\partial_{\hat\tau} X^i [(\partial_{\hat\sigma} X)^2 + B^2]-\partial_{\hat\sigma} X^i [(\partial_{\hat\tau} X \partial_{\hat\sigma} X)-EB]\over \mathcal{L}}\Bigr]~~~~~~~~~~~~~~~~~~~~~~\nn
&~&~~~~~~~~~~~~~~+\partial_{\hat\sigma}\Bigl[{\partial_{\hat\sigma} X^i [(\partial_{\hat\tau} X)^2 + E^2-b^2]-\partial_{\hat\tau} X^i [(\partial_{\hat\tau} X \partial_{\hat\sigma} X)-EB]\over \mathcal{L}}\Bigr]=0
\label{eq1}
\eea
\be
\partial_{\hat\tau}\Bigl[{E (\partial_{\hat\sigma} X)^2+B(\partial_{\hat\tau} X\partial_{\hat\sigma} X)\over \mathcal{L}}\Bigr]-
\partial_{\hat\sigma}\Bigl[{B [(\partial_{\hat\tau} X)^2-b^2]+E(\partial_{\hat\tau} X\partial_{\hat\sigma} X)\over \mathcal{L}}\Bigr]=0
\nonumber
\label{eq2}
\ee
To verify that these equations are satisfied by the configuration (\ref{fullsolexact}) we need following 
identities:
\bea
-\frac{\mathcal{L}}{T_2}&\!\!=\!\!&
\Bigl[{b^4\over 2}+4(\partial_+X_+)^2(\partial_-X_-)^2+4(\partial_+X_+ \partial_- X_-)^2\nonumber\\
&&\,\,+ b^2 [2 S_+ S_- - 2 (\partial_+X_+ \partial_- X_-)-(\partial_+X_+)^2-(\partial_-X_-)^2 ]-8 S_+ S_- 
(\partial_+X_+ \partial_- X_-)\Bigr]^{1/2}\nonumber\\
&\!\!=\!\!&{b^2\over 2}+2 S_+ S_- -2(\partial_+X_+ \partial_- X_-)
\eea
\bea
(\partial_{\hat\sigma} X)^2 + B^2&=&-[(\partial_{\hat\tau} X)^2 + E^2-b^2]=
{b^2\over 2}+2 S_+ S_- -2(\partial_+X_+ \partial_- X_-)\nn
(\partial_{\hat\tau} X \partial_{\hat\sigma} X)-EB&=&0\nn
E (\partial_{\hat\sigma} X)^2+B(\partial_{\hat\tau} X\partial_{\hat\sigma} X)&\!\!=\!\!&2S_+[(\partial_- X_-)^2-(\partial_+X_+\partial_-X_-)]\nonumber\\
&\!\!-\!\!&2S_-[(\partial_+ X_+)^2-(\partial_+X_+\partial_-X_-)]\nonumber\\
&\!\!=\!\!&(S_+-S_-)\,\Bigl[{b^2\over 2}+2 S_+ S_- -2(\partial_+X_+ \partial_- X_-)\Bigr]\nn
B [(\partial_{\hat\tau} X)^2-b^2]+E(\partial_{\hat\tau} X\partial_{\hat\sigma} X)&\!\!=\!\!&
2S_+\Bigl[{b^2\over 2}-(\partial_- X_-)^2-(\partial_+X_+\partial_-X_-)\Bigr]\nonumber\\
&\!\!-\!\!&2S_-\Bigl[-{b^2\over 2}+(\partial_+ X_+)^2+(\partial_+X_+\partial_-X_-)\Bigr]\nonumber\\
&\!\!=\!\!&(S_++S_-)\,\Bigl[{b^2\over 2}+2 S_+ S_- -2(\partial_+X_+ \partial_- X_-)\Bigr]
\eea
Then the  equations (\ref{eq1}),(\ref{eq2}) become
\bea
&&\partial_{\hat\tau}^2 X^i -\partial_{\hat\sigma}^2 X^i = 4 \partial_+ \partial_- X^i = 0\nonumber\\
&&\partial_{\hat\tau} (S_+-S_-)-\partial_{\hat\sigma} (S_++S_-)=2 \partial_- S_+ - 2 \partial_+ S_-=0
\eea
which are seen to be satisfied due the harmonic nature of the fields $X^i,y$.

\subsection{`Quasi-oscillations'}

In the introduction we termed the periodic behavior of the supertube a `quasi-oscillation'. In a regular `oscillation' there is an equilibrium point; if we displace the system from this point then there is a force tending to restore the system to the equilibrium point. But in the supertube we can displace a stationary configuration to a nearby stationary configuration, 
and the system does not try to return to the first configuration. The only time we have such a behavior for a usual {\it oscillatory} system is when we have a `zero mode' (\ref{zero}). Such zero modes allow a `drift' in which  we give the system  
a small initial velocity and then we have an evolution like (\ref{drift}). But the supertube does not have this behavior either; there is no `drift'.\footnote{By contrast, `giant gravitons' have usual vibration modes \cite{giant}. The giant graviton in $AdS_3\times S^3$ has a zero mode corresponding to changing the radius of the giant graviton, and we find  a `drift' over the values of this radius. In \cite{lm6} giant gravitons were studied for $AdS_3$ and it was argued that they give  {\it unbound} states where one brane is separated from the rest \cite{lm6, tachyon}.} 

Now  consider a different system, a particle with charge $e$ and mass $m$ placed in a uniform magnetic field $F_{xy}=B$. 
With the gauge potential $A_y=x$ we have the lagrangian
\be
L={m\over 2}[(\dot x)^2+(\dot y)^2]+e\vec A\cdot \vec v={m\over 2}[(\dot x)^2+(\dot y)^2]+e\, x\dot y
\label{lag3}
\ee
The equations of motion are
\be
\ddot x={e\over m}\dot y, ~~~~\ddot y=-{e\over m}\dot x
\ee
Since each term in the equation has at least one time derivative, any constant position $x=x_0,\, y=y_0$ is a solution. But if we perturb the particle slightly then the particle does not drift over this space of configurations in the manner (\ref{drift}); instead it describes a circle with characteristics (\ref{quasi}). So while this motion is periodic the physics is not that of usual oscillations, and we call it a `quasi-oscillation'. 

Now we wish to show that the motion of the  supertube is also a `quasi-oscillation'.  We will take a simple configuration of the D2 brane to illustrate the point.
Let the D2 brane  extend along  the $z-y$ plane and oscillate in one transverse 
direction $x$. We will restrict to motions which are invariant in $y$ and thus
described by a field $x=x(t,z)$. We will also turn on a ($y$-independent) world volume gauge 
field, for which we choose the $A_t=0$ gauge:
\bea
&&A=A_z(t,z)\,dz+A_y(t,z)\,dy\\
&&F=\dot{A}_z\,dt\wedge dz+\dot{A}_y\,dt\wedge dy + A'_y\,dz\wedge dy\equiv 
\dot{A}_z\,dt\wedge dz+E\,dt\wedge dy - B\,dz\wedge dy\nonumber
\eea
Using $t$, $z$ and $y$ as world volume coordinates, the DBI lagrangian density is
\bea
{\mathcal{L}\over T_2}=-\sqrt{-\mathrm{det}(g+F)}=-[1-\dot{x}^2+x'^2+B^2 (1-\dot{x}^2)-E^2 (1+x'^2)-2 E B\,\dot{x}x'-\dot{A}_z^2]^{1/2}\nonumber\\
\eea
In order to have a qualitative understanding of the dynamics induced by this lagrangian, let us
expand it around a classical stationary solution with $x=0$, ${\bar E}=1$, $B={\bar B}$ and $A_z=0$. We denote by
$a_y(t,z)$ the fluctuation of the gauge field  $A_y$, so that
\be
E=1+\dot{a}_y\,,\quad B={\bar B}-a'_y
\ee
As the gauge field $A_z$ decouples from all other fields we will set it to zero.
Keeping terms up to second order in $x$ and $a_y$, we find the quadratic lagrangian density to be
\be
L^{(2)}=-{\bar B}+{\dot{a}_y\over {\bar B}}+a'_y+{1+{\bar B}^2\over 2 {\bar B}}\,\dot{x}^2+\dot{x}\,x'+
{1\over {\bar B}^2}\,\Bigl({1+{\bar B}^2\over 2 {\bar B}}\,\dot{a}_y^2+\dot{a}_y\,a'_y\Bigr)
\ee 
The terms of first order in $a_y$ are total derivatives (with respect to $t$ and $z$) and do
not contribute to the action. The fields $x$ and $a_y$ are decoupled, at this order,
and both have a lagrangian of the form
\be
L^{(2)}_\phi={1+{\bar B}^2\over 2 {\bar B}}\,\dot{\phi}^2+\dot{\phi}\,\phi'
\label{lag2}
\ee
(with $\phi=x$ or $a_y$). As we can see the lagrangian (\ref{lag2}) has no potential terms (terms
independent of $\dot{\phi}$) and we find that any time independent configuration solves the equations of motion. There is however a magnetic-type interaction
($\dot{\phi}\,\phi'$), which is responsible for the fact that all time dependent solutions are 
oscillatory. Indeed, the equations of motion for $\phi$ are
\be
{1+{\bar B}^2\over 2 {\bar B}}\,\ddot{\phi}+\dot{\phi}'=0
\ee 
whose solution is
\be
\phi=e^{ik\,z -i\omega\,t}\,\,,\,\,~~~~~~\omega= 2\,{\bar B\over 1+{\bar B}^2}\,k
\ee
One can make the analogy between the interaction $\dot{\phi}\,\phi'$ and the toy problem of a particle in a magnetic field more precise by discretizing the $z$ direction on a lattice of spacing $a$. Then we have
\bea
\int dz\, L^{(2)}_\phi &\approx& a\,\sum_n\,\Bigl({m\over 2}\,
 \dot{\phi}_n^2+ \dot{\phi}_n\,{\phi_{n+1}-\phi_{n}\over a}\Bigr)
\nonumber\\
 &\approx& a\,\sum_n\,\Bigl({m\over 2}\,\dot{\phi}_n^2+ {\dot{\phi}_n\,\phi_{n+1}\over a}\Bigr)
\eea
where in the second line we have discarded a total time-derivative and 
$m=(1+{\bar B}^2)/{\bar B}$. The term 
$\dot{\phi}_n\,\phi_{n+1}$ is just like the term $x\dot y$ in (\ref{lag3}) induced by a constant magnetic field
where the variables  $\phi_n,\phi_{n+1}$ play the role of $x,y$.

\subsection{Summary}

We have obtained the full dynamics of the D2 supertube, by mapping the problem to a free string which can be exactly solved. In the D2 language it is not obvious that the problem separates into a `right mover' and a `left mover', but 
(\ref{fullsolexact}) exhibits such a break up. This breakup needs a world sheet coordinate 
$\hat\sigma$ that is a conformal coordinate on the string world sheet, and is thus not an obvious coordinate in the D2 language. The D2 has a natural parametrization in terms of the angular coordinate on the spacetime plane $(X_1,X_2)$, and the difficulties we encountered in mapping the NS1-P solution to the D2 supertube all arose from the change of parametrization. 

\section{The thin tube limit of the gravity solution}\label{grav}
\setcounter{equation}{0}

\begin{figure}
\vskip -.3truein
\begin{center}
\hskip -1.0truein \includegraphics[width=4.5in,angle=270]{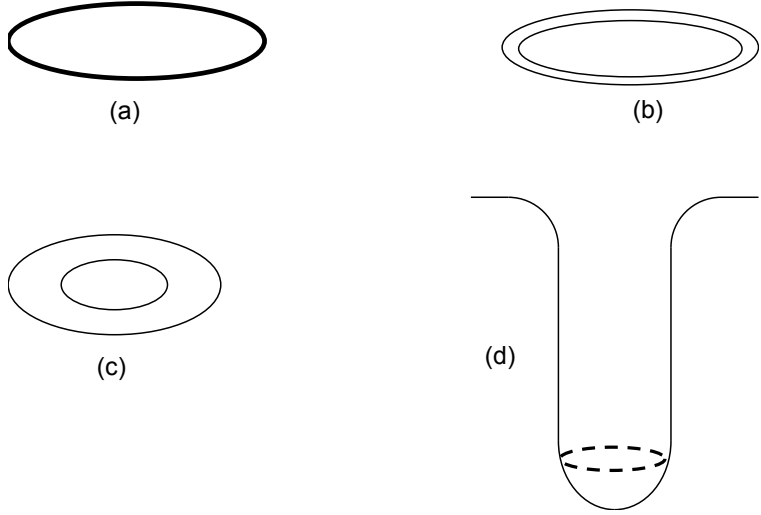}
\end{center}
\vskip -1.0 truein
\caption[x] {{\small (a) The supertube at $g\r 0$, described by a worldsheet action.
(b) The `thin tube' at weak coupling. (c) The `thick tube'
reached at larger coupling. (d) At still larger coupling we get a `deep throat' geometry; the
strands of the NS1 generating the geometry run along the dotted curve.}}
\label{fig2m}
\end{figure}

So far we have ignored gravity in our discussion of the supertube, so we were at vanishing coupling $g= 0$. If we slightly increase $g$ then  the gravitational field of the supertube will extend to some distance off the tube, but for small enough $g$ 
this distance will be much less than the radius of the supertube. We will call this the `thin tube limit', and we picture it in Fig.\ref{fig2m}(b). 

We expect that in this thin tube the dynamics should not be too different from that found at $g\r 0$, and we will find that such is the case; we will find periodic excitations with frequency agreeing with that found from the free string computation and the D2 brane DBI action. But by doing the problem in a gravity description we move from the worldsheet theory to a spacetime one, which will help us to understand what happens when we increase the coupling still further.

Let us recall the 2-charge BPS geometries made in the NS1-P duality frame \cite{lm4}.  Start with type IIB string theory and take the compactification $M_{9,1}\r M_{4,1}\times S^1\times T^4$. As before the coordinate along $S^1$ is $y$ and the coordinates $z_a, a=1\dots 4$ are the coordinates on $T^4$.  The $S^1$ has length $L_y=2\pi R_y$ and the $T^4$ has volume $(2\pi)^4\,V$. The four noncompact spatial directions are called ${\bar x}_i, i=1\dots 4$.  We also write $u=t+y, \, v=t-y$. 

The NS1 is wrapped $n_1$ times around the $S^1$, and carries $n_p$ units of momentum along the $S^1$. This momentum is carried by transverse traveling waves; we assume that the polarization of the wave is in the four noncompact directions and is described by a function $\vec F(v)$. Then the string frame metric, B-field and dilaton are
\bea
ds^2_{string}&=&{H}^{-1}[-du dv +{K}\,dv^2+2 {A}_i\,dv\,d{\bar x}_i]+
d{\bar x}_i d{\bar x}_i + dz_a dz_a\nonumber\\
B&=&{{H}^{-1}-1\over 2}\,du\wedge dv+{H}^{-1}\,{A}_i\,dv\wedge d{\bar x}_i\nonumber\\
e^{2\Phi}&=&{H}^{-1}
\label{fp}
\eea
with
\bea
{H}&=&1+{{\bar Q}_1\over L_T}\,\int_0^{L_T} {dv\over \sum_i({\bar x}_i -F_i(v))^2}\nonumber\\
{K}&=&{{\bar Q}_1\over L_T}\,\int_0^{L_T} dv\,{\sum_i(\dot{F}_i(v))^2\over 
\sum_i({\bar x}_i -F_i(v))^2}\nonumber\\
{A}_i&=&-{{\bar Q}_1\over L_T}\,\int_0^{L_T} dv\,{\dot{F}_i(v)\over 
\sum_i({\bar x}_i -F_i(v))^2}
\label{fpbis}
\eea
Here $L_T=2\pi\,n_1\,R_y$ is the total length of the multiply wound string.

The points on the NS1 spread out over a region in the noncompact directions with size of order $\sim |\vec F(v)|$. On the other hand the gravitational field of the NS1-P system is characterized by the length scales $(\bar Q_1)^{1/2}, \,  (\bar Q_p)^{1/2}$ where
\be
{\bar Q}_p={{\bar Q}_1\over L_T}\,\int_0^{L_T} dv\,\sum_i(\dot{F}_i(v))^2
\label{qbarp}
\ee
In terms of microscopic quantities we have
\be
{\bar Q}_1={g^2 \alpha'^3\over V}\,n_1\,,\quad {\bar Q}_p={g^2 \alpha'^4\over V R_y^2}\,n_p
\label{integercharges}
\ee
Thus when we keep other parameters fixed and take $g$ very small then the gravitational field of the supertube gets confined to a small neighborhood of the supertube and we get a `thin tube' like that pictured in Fig.\ref{fig2m}(b).  If we increase $g$ large then we pass to a `thick tube' like 
Fig.\ref{fig2m}(c) and then to the `deep throat' geometry of Fig.\ref{fig2m}(d).  We can thus say that Fig.\ref{fig2m}(a) is `weak coupling' and 
Fig.\ref{fig2m}(d) is `strong coupling' but note that for `strong coupling' $g$ itself does not need to be large since the charges $n_1, n_p$ are large in (\ref{integercharges}). Thus to be more correct we should say that Fig.2(d) is obtained for large `effective' coupling.

In this section we will consider the `weak coupling' case so that we have a `thin tube'. Then to study the nontrivial part of the metric we have to go close to a point on the tube, so the tube looks essentially like an infinite straight line. Let $z$ be a coordinate along this line (not to be confused with
$z_a$, which are coordinates on $T^4$) and $r$ the radial coordinate
for the three-space perpendicular to the ring. The NS1-P profile was described by a function $\vec F(v)$; let $v=v_0$ correspond to the point $z=0$ along the ring and choose the orientation of the $z$ line such that $z$ increases when $y$
increases. Then we have
\be
z\approx -|\dot{\vec{F}}(v_0)|(v-v_0)\,,\quad \sum_i({\bar x}_i -F_i(v))^2\approx z^2+r^2
\ee
Since we are looking at distances $r$ from the ring which are much smaller than the size of the ring we have
\be
r\ll |\vec{F}(v_0)|
\label{near}
\ee
We can thus make the following approximations
\bea
{H}&\approx&1+{{\bar Q}_1\over L_T\, |\dot{\vec{F}}(v_0)|}\,\int_{-\infty}^{\infty} 
{dz\over z^2+r^2}= 1+{{\bar Q}_1\,\pi\over L_T\, |\dot{\vec{F}}(v_0)|}\,{1\over r}\nonumber\\
{K}&\approx&   {{\bar Q}_1\,|\dot{\vec{F}}(v_0)|\,\pi\over L_T}\,{1\over r}\,,\quad
{A}_z\approx {{\bar Q}_1\,\pi\over L_T}\,{1\over r}
\eea
Define the charge densities
\be
Q_1\equiv {{\bar Q}_1\,\pi\over L_T\, |\dot{\vec{F}}(v_0)|}\,,\quad 
Q_p\equiv {{\bar Q}_1\,|\dot{\vec{F}}(v_0)|\,\pi\over L_T}
\label{qqbar}
\ee
Then we get the geometry  (in the string
frame)
\bea
ds^2_{string}&=&H^{-1}\,[-2 dt\,dv+ \Kt\,dv^2+2 A\,dv\,dz]+dz^2+dx_i dx_i+dz_a dz_a\nonumber\\
B&=&(H^{-1}-1)\,dt\wedge dv+H^{-1}\,A\,dv\wedge dz\nonumber\\
e^{2\Phi}&=&H^{-1}
\label{fpnear}
\eea
\be
H=1+{Q_1\over r}\,,\quad \Kt=1+K=1+{Q_p\over r}\,,\quad A={\sqrt{Q_1\,Q_p}\over r}
\ee
Here we use $x_i$, $i=1,2,3$, to denote the three spatial noncompact directions transverse to the tube.

We are looking for a perturbation of (\ref{fpnear}) corresponding to a deformation of the string profile.
The profile could be deformed either in the
non-compact $x_i$ directions or in the $T^4$ directions. We consider deformations in one of the directions of the $T^4$; this maintains symmetry around the tube in the noncompact directions and is thus easier to work with. 
We thus consider deforming the string profile in one of the $T^4$ directions, denoted $\bar a$. We will
also smear the perturbed metric on $T^4$, so that our fields will be independent on $z_a$.
 
The BPS geometry (\ref{fpnear}) carries a wave of a definite chirality: let us call it right moving. 
If the deformation we add also corresponds to a right moving wave, the resulting geometry can be generated by Garfinkle-Vachaspati transform \cite{gv,lmm}. This will alter the
metric and B-field as follows:
\bea
ds^2_{string} \to ds^2_{string} +2\,\mathcal{A}^{(1)}\,dz_{\bar a}\,,\quad B\to B + \mathcal{A}^{(2)}\wedge dz_{\bar a} 
\label{pertform}
\eea
where
\be
\mathcal{A}^{(1)}=\mathcal{A}^{(2)}= H^{-1}\,a_v\,dv
\label{bpspert}
\ee
and $a_v$ is a harmonic function on  $\mathbb{R}^3\times S^1_z$, whose form will be given in section 
(\ref{secaplus}). 
If we also add a left moving deformation, thus breaking the BPS nature of the system, we do not have a way 
to generate the solution. Note, however, that the unperturbed system has a symmetry under
\be
z_{\bar a}\to -z_{\bar a}
\label{parity}
\ee
 and the perturbation will
be odd under such transformation. We thus expect that only the components of the metric and B-field 
which are odd under (\ref{parity}) will be modified at first order in the perturbation. We can thus
still write the perturbation in the form (\ref{pertform}), with $\mathcal{A}^{(1)}$ and $\mathcal{A}^{(2)}$
some gauge fields on $\mathbb{R}^{(3,1)}\times S^1_z\times S^1_y$, 
not necessarily given by (\ref{bpspert}).

To find the equations of motion for $\mathcal{A}^{(1)}$ and $\mathcal{A}^{(2)}$ we look at the theory dimensionally reduced 
on $T^4$, using the results of \cite{ms}. At first order in the perturbation the dimensionally reduced metric
$g_6$ is simply given by the six-dimensional part of the unperturbed metric (\ref{fpnear}). 
The part of the action involving the gauge fields is  
\be
S_\mathcal{A}=\int\sqrt{-g_6}\,e^{-2\Phi}\,\Bigl[-{1\over 4}\,(F^{(1)})^2 -{1\over 4}\,(F^{(2)})^2-{1\over 12}\,{\tilde H}^2\Bigr]
\ee
where all the index contractions are done with $g_6$. $F^{(1)}$ and $F^{(2)}$ are the usual field strengths
of $\mathcal{A}^{(1)}$ and $\mathcal{A}^{(2)}$ while the field strength $\tilde H$ of 
the dimensionally reduced B-field 
${\tilde B}$ includes the following Chern-Simons couplings:
\bea
&&{\tilde H}_{\mu\nu\lambda}=\partial_\mu {\tilde B}_{\nu\lambda}-{1\over 2}(\mathcal{A}^{(1)}_\mu\,F^{(2)}_{\nu\lambda}+ \mathcal{A}^{(2)}_\mu\,F^{(1)}_{\nu\lambda})+{\rm cyc.~perm.}\nonumber\\
&&{\tilde B}_{\mu\nu}=B_{\mu\nu}+{1\over 2}(\mathcal{A}^{(1)}_\mu\,\mathcal{A}^{(2)}_\nu- \mathcal{A}^{(2)}_\mu\,\mathcal{A}^{(1)}_\nu)
\eea
Using $\tilde B$, $\mathcal{A}^{(1)}$ and $\mathcal{A}^{(2)}$ as independent fields, we find that 
the linearized equations of motion for the gauge fields are
\bea
\nabla^\mu(e^{-2\Phi}\,F^{(1)}_{\mu\lambda})+{1\over 2}e^{-2\Phi}\,
{H^{\mu\nu}}_\lambda\,F^{(2)}_{\mu\nu}=0\,,\quad \nabla^\mu(e^{-2\Phi}\,F^{(2)}_{\mu\lambda})+{1\over 2}e^{-2\Phi}\,
{H^{\mu\nu}}_\lambda\,F^{(1)}_{\mu\nu}=0
\eea
These can be rewritten as decoupled equations as
\be
\mathcal{A}^{\pm}=\mathcal{A}^{(1)}\pm \mathcal{A}^{(2)}
\ee
\be
\nabla^\mu(e^{-2\Phi}\,F^{\pm}_{\mu\lambda})\pm {1\over 2}e^{-2\Phi}\,
{H^{\mu\nu}}_\lambda\,F^{\pm}_{\mu\nu}=0
\label{eqa}
\ee
Our task is to find the solutions of these equations representing non-BPS 
oscillations of the two charge system (\ref{fpnear}). 
 
\subsection{Solution in the `infinite wavelength limit'}\label{solu}

The geometry (\ref{fp}) has a singularity at the curve $\vec x=\vec F(v)$, which is the location of the strands of the oscillating NS1. Since we wish to add perturbations to this geometry, we must understand what boundary conditions to impose at this curve. The wavelength of the oscillations will be of order the length of the tube. Since the tube is `thin' and we look close to the tube, locally the tube will look like a straight line even after the perturbing wave is added. The wave can `tilt' the tube, and give it a velocity. So in this subsection we write the metric for a straight tube which has been rotated and boosted  by infinitesimal parameters $\alpha, \beta$. In the next subsection we will require that close to the axis of the tube (where the singularity lies) all fields match onto such a rotated and boosted straight tube solution. 

We will consider oscillations of the supertube in one of the $T^4$ directions. Since we smear on the $T^4$ directions, the solution will remain independent of the torus coordinates $z_a$ but we will get components in the metric and $B$ field which reflect the `tilt' of the supertube. We are using the NS1-P description. The unperturbed configuration looks, locally, like a NS1 that is a slanted line in the $y-z$ plane, where $z$ is the coordinate along the tube. The perturbation tilts the tube towards a 
$T^4$ direction $z_{\bar a}$. We will find it convenient to start with the NS1 along $y$, first add the tilt and boost corresponding to the perturbation, and then add  the non-infinitesimal tilt in the $y-z$ plane (and the corresponding boost).

We start from the one charge system 
\bea
ds^2_{string}&=&H^{-1}\,[-(d{\tilde t}'')^2+(d{\tilde y}'')^2]+
(d{\tilde z}'')^2+dx_i dx_i+
d{\tilde z}_{\bar a}'' d{\tilde z}_{\bar a}''
+\sum_{a\not={\bar a}} dz_a dz_a\nonumber\\
B&=&-(H^{-1}-1)\,d{\tilde t}''\wedge d{\tilde y}''\nonumber\\
e^{2\Phi}&=&H^{-1}
\eea
and perform the following operations: An infinitesimal boost in the direction 
${\tilde z}_{\bar a}''$, 
with parameter $\beta$
\be
{\tilde t}''= {\tilde t}' -  {\tilde z}_{\bar a}'\,\beta\,,\quad 
{\tilde z}_{\bar a}''={\tilde z}_{\bar a}' - {\tilde t}'\,\beta\,,\quad 
{\tilde y}''={\tilde y}'\,,\quad {\tilde z}''={\tilde z}'
\ee
and an infinitesimal rotation in the
$({\tilde y}',{\tilde z}'_{\bar a})$ plane, with parameter $\alpha$:
\be
{\tilde y}'={\tilde y} +{\tilde z}_{\bar a}\,\alpha\,,\quad 
{\tilde z}'_{\bar a}={\tilde z}_{\bar a}-{\tilde y}\,\alpha\,,\quad {\tilde t}'={\tilde t}\,,\quad {\tilde z}'={\tilde z}
\ee
These operations give
\bea
ds^2_{string}
&=&H^{-1}\,\Bigl[-d{\tilde t}^2+d{\tilde y}^2 -2 \alpha\,(H-1)\,d{\tilde y}\,d{\tilde z}_{\bar a} -2 \beta\,(H-1)\,d{\tilde t}\,d{\tilde z}_{\bar a}\Bigr]\nonumber\\
&+&d{\tilde z}^2+dx_i dx_i+d{\tilde z}_{\bar a} d{\tilde z}_{\bar a}+\sum_{a\not={\bar a}} dz_a dz_a\nonumber\\
B
&=&-(H^{-1}-1)\,d{\tilde t}\wedge d{\tilde y}+\beta\,H^{-1}\,(H-1)\,d{\tilde y}\wedge d{\tilde z}_{\bar a}+\alpha\,H^{-1}\,(H-1)\,d{\tilde t}\wedge d{\tilde z}_{\bar a}\nonumber\\
e^{2\Phi}&=&H^{-1}
\label{pertstr}
\eea
We can read off from (\ref{pertstr}) the gauge fields $\mathcal{A}^{\pm}$:
\bea
&&\mathcal{A}^{+}=(\alpha-\beta)\,H^{-1}\,(H-1)\,d{\tilde v}\,,\quad
\mathcal{A}^{-}=-(\alpha+\beta)\,H^{-1}\,(H-1)\,d{\tilde u}
\label{straighttilde}
\eea
($\tilde u={\tilde t}+{\tilde y}$ and $\tilde v={\tilde t}-{\tilde y}$).
The part of the perturbation proportional to $\alpha-\beta$ represents a right moving wave, in which case
only the $\mathcal{A}^{+}$ gauge field is excited. The reverse happens for 
the left moving perturbation, proportional to $\alpha+\beta$. 

We would now like to   add a finite amount of momentum $Q_p$ to the system (\ref{pertstr}). This momentum
is carried by a right moving wave moving with the speed of light in the positive $y$ direction, with polarization  in the direction $z$. 
The result will give us a geometry representing
a small perturbation of the system (\ref{fpnear}). We can reach the desired configuration from 
(\ref{pertstr}) by
 performing
a boost in the direction $\tilde z$ with parameter ${\bar \beta}$
\be
{\tilde t}= t'\, \cosh{\bar \beta} -  z'\,\sinh{\bar \beta}\,,\quad 
{\tilde z}= z'\,\cosh{\bar \beta} - t'\,\sinh{\bar \beta}\,,\quad {\tilde y}=y'\,,\quad {\tilde z}_{\bar a}=z'_{\bar a}
\ee
followed by a rotation in the
$(y',z')$ plane, with parameter ${\bar \alpha}$:
\be
y'=y\,\cos{\bar \alpha}+z\,\sin{\bar \alpha}\quad 
z'=z\,\cos{\bar \alpha}-y\,\sin{\bar \alpha}\,,\quad t'=t\,,\quad 
z'_{\bar a}=z_{\bar a}
\ee

\begin{figure}
\vskip -.5truein
\begin{center}
\hskip -.5truein \includegraphics[width=4.in,angle=270]{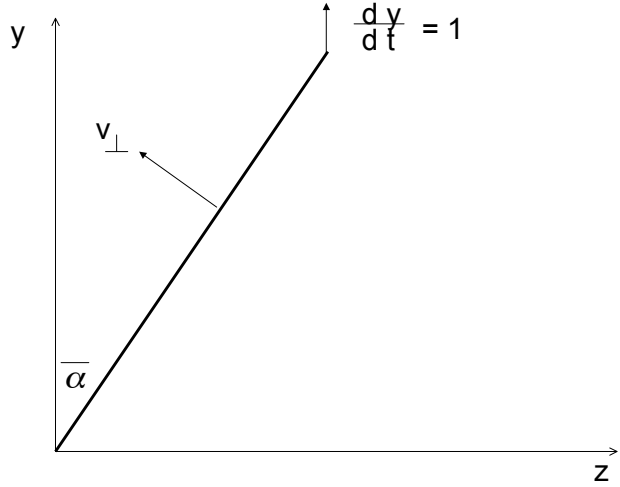}
\end{center}
\vskip -1.0truein
\caption[x] {{\small A short segment of the NS1 moving at the 
speed of light in the $y$ direction. This yields a velocity $v$ for the
 segment in the direction perpendicular to itself.}}
\label{fig3m}
\end{figure}

The parameters $\bar\alpha, \bar\beta$ are related. This is because the segment of string under consideration is supposed to be a short piece of the string in a state like that in Fig.\ref{fig1m}(a), where the traveling wave is moving in the positive $y$ direction with the speed of light. We depict this segment in 
Fig.\ref{fig3m}. We can ask how fast the string segment must be moving in a direction {\it perpendicular} to itself to yield $dy/dt=1$, and we find
\be
v_\perp\equiv-\tanh\bar\beta=\sin\bar\alpha
\ee
This implies
\be
\sinh{\bar \beta}=-\tan{\bar \alpha}\,,\quad \cosh{\bar \beta}={1\over \cos{\bar \alpha}}
\label{ab}
\ee
The final configuration is given by
\bea
ds^2_{string}&=&H^{-1}\,\Bigl[-2 dt\,dv+[1+\sinh^2{\bar \beta}\,(H-1)]\,dv^2- 2\sinh{\bar \beta}\,(H-1)\,dv\,dz \nonumber\\
&+&2 (H-1)\,\Bigl(
{\alpha\,\cos^2{\bar \alpha}-\beta\,\sin^2{\bar \alpha}\over \cos{\bar \alpha}}\,dv\,dz_{\bar a}- 
(\alpha+\beta)\,
\cos{\bar \alpha}\,dt\,dz_{\bar a}\nonumber\\
&-&(\alpha+\beta)\,\sin{\bar \alpha}\,dz\,dz_{\bar a}\Bigr)\Bigr]
+dz^2+dx_i dx_i+dz_{\bar a} dz_{\bar a}+\sum_{a\not={\bar a}} dz_a dz_a\nonumber\\
B&=&(H^{-1}-1)\,dt\,dv-H^{-1}\,(H-1)\,\sinh{\bar \beta}\,dv\wedge dz\nonumber\\
&+&H^{-1}\,(H-1)\Bigl(
{\alpha\,\sin^2{\bar \alpha}-\beta\,\cos^2{\bar \alpha}\over \cos{\bar \alpha}}\,dv\wedge dz_{\bar a}+(\alpha+\beta)\,\cos{\bar \alpha}\,dt\wedge dz_{\bar a}\nonumber\\
&+&(\alpha+\beta)\,\sin{\bar \alpha}\,dz\wedge dz_{\bar a} \Bigr)\nonumber\\
e^{2\Phi}&=&H^{-1}
\eea
We note that, for $\alpha=\beta=0$, we obtain the system (\ref{fpnear}) with\footnote{With our conventions $\bar\alpha>0$ and $\bar\beta<0$. Thus $\sqrt{Q_p}=-\sqrt{Q_1}\,\sinh{\bar\beta}$.}
\be
Q_p=Q_1\,\sinh^2{\bar \beta}
\label{qp}
\ee
The perturbation is proportional to $\alpha$ and $\beta$ and is encoded in the gauge fields
\bea
&&\mathcal{A}^{+}_v=({\tilde\alpha}-{\tilde\beta})\,H^{-1}\,{Q_1\over r}\,,\quad \mathcal{A}^{+}_t=0\,,\quad \mathcal{A}^{+}_z=0\nonumber\\
&&\mathcal{A}^{-}_v=({\tilde\alpha}+{\tilde\beta})\,H^{-1}\,{Q_1-Q_p\over r}\,,
\quad \mathcal{A}^{-}_t=-2({\tilde\alpha}+{\tilde\beta})\,H^{-1}\,{Q_1\over r}\nonumber\\
&&\mathcal{A}^{-}_z=-
2({\tilde\alpha}+{\tilde\beta})\,H^{-1}\,{\sqrt{Q_1 Q_p}\over r}
\label{straight}
\eea  
where we have redefined
\be
{\tilde\alpha}-{\tilde\beta}={\alpha-\beta\over \cos{\bar \alpha}}\,,\quad {\tilde\alpha}+{\tilde\beta}=
(\alpha+\beta)\,\cos{\bar \alpha}
\ee
and we have used (\ref{qp}) and (\ref{ab}). We see that, as before, $\mathcal{A}^{+}$ comes from right moving
perturbations, proportional to ${\tilde\alpha}-{\tilde\beta}$, and $\mathcal{A}^{-}$ comes from left moving
perturbations, proportional to ${\tilde\alpha}+{\tilde\beta}$.

\subsection{Solution for $\mathcal{A}^{+}$}\label{secaplus}

Let us look for a solution of (\ref{eqa}), in the $\mathcal{A}^{+}$ sector,
 which matches the configuration (\ref{straight}) when
$k\to0$. We learned from (\ref{straight}) that $\mathcal{A}^{+}$ receives contributions only from the BPS (right moving) part of the wave and that, at least in the long wavelength limit, only the component  $\mathcal{A}^{+}_v$ is non-vanishing.
One can thus look for a solution of the form
 \be
\mathcal{A}^{+}_v=H^{-1}\,a^{+}_v\,,\quad \mathcal{A}^{+}_t=0\,,\quad  \mathcal{A}^{+}_z=0\,,\quad 
 \mathcal{A}^{+}_i=0
\ee
Equation (\ref{eqa}) implies the following conditions for $a^{+}_v$ (here $\triangle=\p_i\p_i$ is the ordinary  Laplacian in the 3-dimensional space of the $x_i$)
\bea
&&\lambda=t:\quad \partial_t^2 a^{+}_v =0\\
&&\lambda=v:\quad \triangle\, a^{+}_v+\partial_z^2 a^{+}_v- 2A\,
\partial_t\partial_z a^{+}_v=0\\
&&\lambda=z:\quad \partial_t\partial_z a^{+}_v =0\\
&&\lambda=i:\quad \partial_t\partial_i a^{+}_v=0
\eea
It is thus clear that a $t$-independent $a^{+}_v$ satisfying
\be
\triangle\, a^{+}_v+\partial_z^2 a^{+}_v=0
\label{eqa+v}
\ee
solves the linearized equations of motion. The general solution of 
(\ref{eqa+v}), with momentum 
\be
k={n\over R_z}\,\,
\ee 
along $z$, is
\be
a^{+}_v=e^{i k z}\,{c_+\,e^{k\,r}+c_-\,e^{-k\,r}\over r}+{\rm c.c.}
\ee
Without loss of generality let us set $n$ to be positive. To have a converging field at large $r$ one should take $c_+=0$. Matching with (\ref{straight}) fixes $c_-$:
\be
c_-=({\tilde \alpha}-{\tilde \beta})\,Q_1
\ee 
so that
\be
\mathcal{A}^{+}_v=({\tilde \alpha}-{\tilde \beta})\,H^{-1}\,{Q_1\over r}\,e^{i k\, z - k\,r}+{\rm c.c.}
\label{a+sol}
\ee

The above result is consistent with the form of $\mathcal{A}^{+}$ derived by
Garfinkle-Vachaspati transform: Consider a string carrying a right moving
wave described by the profile $F_i(v)$ in the non-compact directions ${\bar x}_i$ and 
$f_{\bar a}(v)$ in the $T^4$ direction $z_{\bar a}$. 
After smearing over $z_{a}$, Garfinkle-Vachaspati transform 
predicts a gauge field
\be
\mathcal{A}^{+}_v=\mathcal{A}_v^{(1)}+\mathcal{A}_v^{(2)}\nonumber
\ee
\be
\mathcal{A}_v^{(1)}=\mathcal{A}_v^{(2)}=-{H}^{-1}\,{{\bar Q}_1\over L_T}\,\int_0^{L_T} dv\,
{\dot{f}_{\bar a}(v)\over \sum_i({\bar x}_i-F_i(v))^2}
\label{vac}
\ee 
Eq.~(\ref{vac}) is analogous to the relation (\ref{fpbis}) for $A_i$, applied to the case
in which the profile extends in the $T^4$ directions. Let us take the near ring limit of (\ref{vac}) 
for a profile $f_{\bar a}$ of the form
\be
f_{\bar a}(v)=\xi_{\bar a}\,e^{-i{\bar k}\,v}+c.c.
\ee
Around some point $v_0$ on the ring we write
\be
z=-|\dot{\vec{F}}(v_0)|\,(v-v_0)\,,\quad z_0=-|\dot{\vec{F}}(v_0)|\,v_0
\ee
so that we can write
\be
f_{\bar a}(v)=\xi_{\bar a}\,e^{i{\bar k}\,(z+z_0)/ |\dot{\vec{F}}(v_0)|}+c.c.
\equiv \xi_{\bar a}\,e^{i k\,(z+z_0)}+c.c.
\ee
and
\be
\dot{f}_{\bar a}(v)= -i\xi_{\bar a}\,|\dot{\vec{F}}(v_0)|\,k\,e^{i k\,(z+z_0)}+c.c.
\ee
In the near ring limit one can approximate
\bea
\mathcal{A}^{+}_v&\approx&2 H^{-1}\,{i {\bar Q}_1\,\xi_{\bar a}\,k\over L_T}\,e^{i k\, z_0}\,
\int_{-\infty}^{+\infty} dz\,{e^{i k\, z}\over r^2+z^2}+{\rm c.c.}
\nonumber\\
&=&2 H^{-1}\,
{i{\bar Q}_1\,\pi\,\xi_{\bar a}\,k\over L_T}\,{e^{i k\, z_0-k\, r}\over r}+{\rm c.c.}
\label{vachbis}
\eea
Using (\ref{qqbar}) to relate ${\bar Q}_1$ and $Q_1$
we see that (\ref{vachbis}) coincides with (\ref{a+sol}), with
\be
({\tilde \alpha}-{\tilde \beta})= 2 i\,\xi_{\bar a}\,|\dot{\vec{F}}(v_0)|\,k
\ee

The time-independent solution (\ref{a+sol}) represents the response of the system to a BPS right moving
wave. Since
the $\mathcal{A}^{+}$ part of the gauge field should only be sensitive to BPS deformations, we expect that 
equation (\ref{eqa}) for $\mathcal{A}^{+}$ should not admit time-dependent solutions consistent with the 
boundary condition (\ref{straight}).   
In an appendix we prove this fact for the more general $\mathcal{A}^{+}$ ansatz.

\subsection{Solution for $\mathcal{A}^{-}$}
\label{aminus}

We now look at the $\mathcal{A}^{-}$ sector, where we expect to find the time-dependent configurations
corresponding to left moving non-BPS perturbations.

Consider an ansatz of the form
\be
\mathcal{A}^{-}_v=H^{-1}\,a^{-}_v\,,\quad \mathcal{A}^{-}_t=H^{-1}\,a^{-}_t\,,\quad  \mathcal{A}^{-}_z=H^{-1}\,a^{-}_z\,,\quad 
 \mathcal{A}^{-}_i=0
\label{minus1}
\ee
By spherical symmetry $\mathcal{A}^-_i$ only has a radial component $\mathcal{A}^-_r$ and we chose our 
gauge to set $\mathcal{A}^-_r=0$. (Such a gauge can  have difficulties at $r=0$ but we can consider it as an ansatz and see later that we obtain a good solution.)
The equations for $a^{-}_v$, $a^{-}_t$ and $a^{-}_z$, obtained by using the ansatz (\ref{minus1})
in (\ref{eqa}) and using the background (\ref{fpnear}), are (we list the equations in the order
$\lambda=t,v,z,i$)
\bea
\triangle\, a^{-}_t+H\,\partial_t^2 a^{-}_v + \partial_z(\partial_z a^{-}_t -\partial_t a^{-}_z)+A\,\partial_t(\partial_z a^{-}_t -\partial_t a^{-}_z)=0
\label{eqt-}
\eea
\bea
&&\triangle\, a^{-}_v + \partial_z^2 a^{-}_v - [(H \Kt -A^2)\,\partial_t^2 a^{-}_v - 2 A\,\partial_t\partial_z a^{-}_v]\nonumber\\
&&+H^{-2}\,\partial_i H\partial_i H\,(2 a^{-}_v + \Kt \,a^{-}_t)-H^{-1}\,\partial_i H\,\partial_i (2 a^{-}_v +\Kt \,a^{-}_t)+
\partial_i a^{-}_t\,\partial_i \Kt=0\nonumber\\
\label{eqv-}
\eea
\bea
&&
\triangle\, a^{-}_z+H\,\partial_t\partial_z a^{-}_v+(H \Kt -A^2)\,\partial_t(\partial_z a^{-}_t -\partial_t a^{-}_z)-A\,\partial_z
(\partial_z a^{-}_t -\partial_t a^{-}_z)\nonumber\\
&&+2H^{-2}\,\partial_i H\partial_i H (a^{-}_z + A\,a^{-}_t)-2 H^{-1}\,\partial_i H\,\partial_i \,(a^{-}_z + A\,a^{-}_t)+2
\partial_i a^{-}_t\,\partial_i A=0\nonumber\\
\label{eqz-}
\eea
\bea
&&H\,\partial_t\partial_i a^{-}_v -\partial_z\partial_i a^{-}_z + [(H \Kt-A^2)\,\partial_t\partial_i a^{-}_t - A\,
\partial_z\partial_i a^{-}_t - A\,\partial_t \partial_i a^{-}_z]\nonumber\\
&&+H^{-1}\,\partial_i H\,[\partial_z a^{-}_z + A\,(\partial_t a^{-}_z + \partial_z a^{-}_t)-(H \Kt-A^2)\,\partial_t a^{-}_t ]\nonumber\\
&&-\partial_i A\,(\partial_z a^{-}_t -\partial_t a^{-}_z)-2\partial_i H\,\partial_t a^{-}_v=0
\label{eqi-}
\eea
Inspired by the limiting solution (\ref{straight}), we make the following ansatz for $a^{-}_v$, $a^{-}_t$ and $a^{-}_z$:
\bea
&&a^{-}_v=(Q_1-Q_p)\,e^{i k\, z- i\omega\, t}\,f(r)\nonumber\\
&&a^{-}_t=-2 Q_1\,e^{i k\, z- i\omega\, t}\,f(r)\,,\quad
a^{-}_z=-2 \sqrt{Q_1 Q_p}\,e^{i k\, z- i\omega\, t}\,f(r)
\label{ansatz-}
\eea
Substituting this ansatz in eq.~(\ref{eqt-}) we find an equation for $f(r)$:
\be
-2 Q_1\,\triangle\, f -\omega^2\,(Q_1-Q_p)\,f + 2 k\,(k Q_1+\omega\,\sqrt{Q_1 Q_p})\,f - 
{\omega\,Q_1\,f\over r}[2\sqrt{Q_1 Q_p}\,k+(Q_1+Q_p)\,\omega]=0
\label{eqt-bis}
\ee 
This equation can be simplified by taking
\be
f={{\tilde f}\over r}
\ee
after which we get
\be
-2 Q_1\,{\tilde f}'' -\omega^2\,(Q_1-Q_p)\,{\tilde f} + 2 k\,(k Q_1+\omega\,\sqrt{Q_1 Q_p})\,{\tilde f} 
- {\omega\,Q_1\,{\tilde f}\over r}[2\sqrt{Q_1 Q_p}\,k+(Q_1+Q_p)\,\omega]=0
\label{eqt-tris}
\ee 
According to the boundary condition  (\ref{straight}), we want $\tilde f$ to go to a constant when 
$r\to 0$; this is only possible if the $1/r$ term in (\ref{eqt-tris}) vanishes and this determines the frequency of oscillation
to be
\be
\omega=-k\,{2\sqrt{Q_1 Q_p}\over Q_1+Q_p}
\label{omega}
\ee
Using this value of $\omega$ back in (\ref{eqt-tris}) we find that $\tilde f$ satisfies
\be
{\tilde f}'' -{\tilde k}^2 {\tilde f} =0
\ee
with
\be
{\tilde k}^2=k^2-\omega^2=k^2\,\Bigl({Q_1-Q_p\over Q_1+Q_p}\Bigr)^2 
\label{ktilde}
\ee
and thus
\be
{\tilde f}=c_+\,e^{+|{\tilde k}|\,r}+ c_-\,e^{-|{\tilde k}|\,r}
\ee
In order to have a converging solution for large $r$ one needs $c_+=0$ and to match with (\ref{straight})
one needs $c_-={\tilde\alpha}+{\tilde\beta}$.
To summarize we find 
\bea
\label{sol-}
&&\mathcal{A}^{-}_v=({\tilde\alpha}+{\tilde\beta})\,H^{-1}\,(Q_1-Q_p)\,e^{i k\, z- i\omega\, t}\,{e^{-|{\tilde k}|\,r}\over r}\\
&&\mathcal{A}^{-}_t=-2 ({\tilde\alpha}+{\tilde\beta})\,H^{-1}\,Q_1\,e^{i k\, z- i\omega\, t}\,{e^{-|{\tilde k}|\,r}\over r}\,,\quad
\mathcal{A}^{-}_z=-2({\tilde\alpha}+{\tilde\beta})\,H^{-1}\sqrt{Q_1 Q_p}\,e^{i k\, z- i\omega\, t}\,{e^{-|{\tilde k}|\,r}\over r}\nonumber
\eea

It is a lengthy but straightforward exercise to verify that (\ref{sol-}) solves the remaining equations 
(\ref{eqv-}-\ref{eqi-}).

\subsection{Period of the oscillations}

The speed of the left-moving wave on the supertube is
\be
v={\omega\over |k|}=2{\sqrt{Q_1 Q_p}\over Q_1+Q_p}
\label{velocity3}
\ee
The direction $z$ used above is the coordinate along the supertube. So even though $z$ looked like an infinite direction
in the `near tube' limit, this direction is actually a closed curve with a length $L_z$. The time for the wave to travel around this closed curve is
\be 
\Delta t=\int_0^{L_z} {dz\over v}=\int_0^{L_z}dz {Q_1+Q_p\over 2\sqrt{Q_1 Q_p}}={1\over 2}
\int_0^{L_z} dz [\sqrt{Q_1\over Q_p}+\sqrt{Q_p\over Q_1}]
\label{periodf}
\ee
We have
\be
Q_1={{\bar Q}_1\,\pi\over L_T}\,{1\over \eta}\,,\quad Q_p={{\bar Q}_1\,\pi\over L_T}\,\eta
\label{qqp}
\ee 
with 
\be
\eta^{-1} ={1\over |\dot{\vec{F}}|} = {dy\over dz}
\label{eta}
\ee
This gives
\bea
\Delta t&=&{1\over 2}\int_0^{L_z} dz (\eta^{-1}+\eta)\nn
&=&{1\over 2}\int_0^{L_z} dz[{dy\over dz}+ {dy\over dz} ({dz\over dy})^2]\nn
&=&{1\over 2T}(Tn_1L_y)+{1\over 2T}(T\int |\dot F|^2dy)\nn
&=& {1\over 2T} (M_{NS1}+M_P)
\eea
where $M_{NS1}$ is the mass contributed by the NS1 charge and $M_P$ is the mass of the momentum charge. We see that this period $\Delta t$ agrees with the period (\ref{periodp}) found from the  NS1-P system at $g=0$.

We offer an intuitive explanation  for  the time period (\ref{periodp}). We have
\be
{Q_1\over Q_p}=\eta^{-2}=({dy\over dz})^2
\ee
Thus we can write (\ref{velocity3}) as
\be
v=2{{dy\over dz}\over 1+({dy\over dz})^2}
\label{vel1}
\ee
Consider a segment of the NS1 before the perturbation is added. In section (\ref{solu}) we had seen, (with the help of  Fig.\ref{fig3m}) that because this  segment represents a wave traveling in the $y$
direction with $dy/dt=1$,  the velocity of this segment {\it perpendicular} to itself was
\be
v_\perp=\sin\bar\alpha ={1\over \sqrt{1+({dy\over dz})^2}}
\ee
So we have a segment of a NS1, moving at a certain velocity transverse to itself. Go to the rest frame of this segment. Then any small perturbation on the segment will move to the right or to the left with speed unity. Consider the perturbation going left. 

Now return to the original reference frame, and look at this perturbation on the segment. The distances along the segment are not affected by the change of frame (since the boost is perpendicular to the segment) but there is a time dilation by a factor $\gamma=1/\sqrt{1-v_\perp^2}$. This means that the perturbation will be seen to be moving along the strand at a speed
\be
v_L=\gamma^{-1}={{dy\over dz}\over \sqrt{1+({dy\over dz})^2}}
\ee
We are interested in the motion of the perturbation in the $z$ direction, so we look at the $z$ component of this velocity
\be
v_{L,z}=v_L\sin\bar\alpha={{dy\over dz}\over 1+({dy\over dz})^2}
\ee
What we actually observe as the wave on the supertube is a deformation moving along the tube, so we wish to measure the progress of the waveform as a function of the coordinate $z$. A given point on our NS1 segment moves in the direction of the velocity $v$, so it moves towards smaller $z$ values at a speed
\be
v_z=v_\perp \cos\bar\alpha={{dy\over dz}\over 1+({dy\over dz})^2}
\ee
Thus if we measure the speed of the left moving perturbation with respect to a a frame where $z$ is fixed then we find the velocity
\be
v^{pert}_L=v_{L,z}+v_z=2{{dy\over dz}\over 1+({dy\over dz})^2}
\ee
which agrees with (\ref{vel1}).

Similarly if we look at the right moving perturbation then we find
\be
v^{pert}_R=-v_{L,z}+v_z=0
\ee
This agrees with the fact that if we add a further right moving wave to the NS1 then we just get another BPS tube configuration, which is stationary and so does not change with time.

\section{Coupling to radiation modes}

The perturbations of the `thin' tube in the `infinite line limit'  is seen to fall off exponentially with the distance
from the tube axis. Note however that if we take the longest wavelengths on the supertube, then the  term $e^{-{|\tilde k| r}}$
is not really significant. For such modes $|\tilde k|\sim 1/a$ where $a$ is the radius of the tube. So $e^{-{|\tilde k| r}}\sim 1$ for $r\ll a$, and for $ r\gtrsim a$ we cannot use the infinite line limit of the thin tube anyway. If however we look at higher wavenumbers on the tube then $|\tilde k|\sim {n/a}$ and then the factor $e^{-{|\tilde k| r}}$ is indeed significant in describing the fall off of the perturbation away from the tube axis.

We now wish to look at the behavior of the perturbation far from the entire supertube, i.e. for distances $\bar r\gg a$. Here we use the symbol $\bar r$ for the radial coordinate in the 4 dimensional noncompact space, to distinguish it from the radial distance $r$ from the tube axis that we used in the last section when looking at the `infinite line limit'. For $\bar r \gg a$ we get flat space. Suppose we were studying a scalar field $\square \Psi=0$ in the supertube geometry. We can write
\be
\Psi=e^{-i\omega\,t}\,\mathcal{R}(\bar r)\,Y^{(l)}(\theta,\phi,\psi)
\ee
If $\omega^2<0$ then we get solutions $\sim e^{\pm |\omega| t}$; these are not allowed because they will not conserve energy. For $\omega^2>0$, we get the behavior (see Appendix (\ref{asym}))
\be
\mathcal{R}={r_{+}\,e^{i \omega\,\bar r}+r_-\,e^{-i \omega\,\bar r}\over 
{\bar r}^{3/2}}(1+O({\bar r}^{-1}))
\ee
This solution describes traveling waves that carry flux to and from spatial infinity.  Thus if we start with an excitation localized near the supertube then the part of its wavefunction that extends to large $\bar r$ will lead to the energy of excitation flowing off to infinity as radiation.

Let us see how significant this effect is for the `thin tube'. Let us set $Q_1, Q_p$ to be of the same order. From (\ref{sol-}) we see that the magnitude of the perturbation  behaves as 
 \be
\mathcal{A}  \sim H^{-1} {Q_1\over r} \sim {Q_1\over r+Q_1}
 \ee
 Thus if the perturbation is order unity at the ring axis then at distances 
$r \gtrsim a$ we will have
 \be
\mathcal{A} \lesssim {Q_1\over a}
 \ee
 But the thin tube limit is precisely the one where the ratio $Q_1/a$ is small, so the part of the wavefunction reaching large $r$ is small. Thus the rate of leakage of energy to the radiation field is small, and the excitations on the `thin tube' will be long lived.
 This is of course consistent with the fact that in the limit $g\r 0$ we can describe the system by just the free string action or the D2 brane DBI action, and here there is {\it no} leakage of energy off the supertube to infinity.
 
 As we keep increasing $g$ we go from the `thin' tube of Fig.\ref{fig2m}(b) to the `thick tube' of Fig.\ref{fig2m}(c). Now $Q_1/a\sim 1$ and the strength of the perturbation reaching the radiation zone is {\it not} small. We thus expect that the energy of excitation will flow off to infinity in a time of order  the oscillation time of the mode. Thus we expect that the oscillations of the supertube become `broad resonances' and cease to be well defined oscillations as we go from Fig.\ref{fig2m}(a) to Fig.\ref{fig2m}(c).

In the above discussion we referred to the excitation as a scalar field, but this is just a toy model; what we have is a 1-form field in 5+1 spacetime. In Appendix (\ref{asym}) we solve the field equations for this 1-form field at infinity, and find again a fall off at infinity that gives  a non-zero flux of energy. We also find the next correction in $1/{\bar r}$, and show how a series expansion in $1/{\bar r}$ may be obtained in general. These corrections do not change the fact that the leading order term carries flux out to infinity. It is important that the first correction to flat space is a potential $\sim 1/{\bar r}^2$ and not $\sim 1/{\bar r}$; this avoids the appearance of a logarithmic correction at infinity. 

It is to be noted that such series solutions in $1/{\bar r}$ are asymptotic expansions rather than series with a nonzero radius of convergence \cite{book}, so these arguments are not a rigorous proof for the absence of infinitely long lived oscillations.
The wave equations for a given $\omega$  are similar in structure to the  Schrodinger equation (in 4+1 dimensions) with a potential $V$
falling to zero at infinity
\be
-\mathcal{R}''+V({\bar r},\omega)\mathcal{R}=\omega^2 \mathcal{R}
\ee
Note that because $\omega^2>0$ our wavefunction would be like a positive energy eigenstate of the Schrodinger equation; i.e.  we need an energy eigenvalue embedded in the continuum spectrum.
For the Schrodinger equation there are several results that exclude such eigenvalues on general grounds \cite{reedsimon}.
The required results come from two kinds of theorems. First we need to know that there is no `potential well with infinitely high walls' near the origin; if there was such a well then we can have a positive energy eigenstate which has no `tail' outside the well. Next, given that there {\it is} a tail outside the well we need to know that the potential falls off to zero fast enough and does not `oscillate' too much; such oscillations of the potential can cause the wavefunction to be back-scattered towards the origin repeatedly and die off too fast to carry a nonzero flux at infinity. We cannot directly apply these results to our problem because our equations are not exactly the Schrodinger equation, but the potential like terms in our equations do not appear to be of the kind that will prevent flux leakage to infinity. 

To summarize, we conjecture that as we increase $g$ to go from Fig.\ref{fig2m} (a) to Fig.\ref{fig2m}(c) the periodic oscillations present at $g\r 0$ merge into the continuum spectrum of bulk supergravity. Thus for $g>0$ the energy of excitation placed on the supertube 
eventually leaks off to infinity, with the rate of leakage increasing as we go from the `thin tube' to the `thick tube'.  

\section{Long lived excitations at large coupling}\label{long}
\setcounter{equation}{0}

Let us increase the coupling still further. Then the supertube geometry becomes like that pictured in Fig.\ref{fig2m}(d) \cite{bal,mm,lm4}. The metric is flat space at infinity, then we have a `neck' region, this leads to a deep `throat' which ends in a `cap' near $r=0$. Supergravity quanta can be trapped in the `throat' bouncing between the cap and the neck for long times before escaping to infinity. We first consider the gravity description, then a microscopic computation, and finally suggest a relation between the two.

\subsection{The geometry at large effective coupling}

Consider an NS1 wrapped $n_1$ times on the $S^1$ with radius $R_y$, and give it the transverse vibration profile 
\be
X_1=a\cos {(t-y)\over n_1 R_y}, ~~~X_2=a\sin {(t-y)\over n_1 R_y}
\label{profile}
\ee
Thus the string describes a `uniform helix with one turn' in the covering space of the $S^1$. At weak coupling $g\r 0$ we get a ring with radius $a$ in flat space, while at strong coupling we get a geometry like Fig.\ref{fig2m}(d) with the circle (\ref{profile}) sinking deep into the throat (the dotted line in the figure).

In \cite{lm4} the computations were done in the D1-D5 duality frame, so let us start with that frame and dualize back to NS1-P later. We will denote quantities in the D1-D5 frame by primes. The time for a supergravity quantum to make one trip down the throat and back up is
\be
\Delta t_{osc}=\pi R'_y
\ee
where $R'_y$ is the radius of $S^1$ in the D1-D5 frame. When the quantum reaches the neck there is a probability $P$ that it would escape to infinity, and a probability $1-P$ that it would reflect back down the throat for another cycle. For low energy quanta in the $l$-th spherical harmonic this probability $P$ is given by 
\cite{maldastrom,lm3}
\be
P_l=4\pi^2({{\bar Q}'_1{\bar Q}'_5\omega'^4\over 16})^{l+1}[{1\over (l+1)!l!}]^2
\ee
where $\omega'$ is the energy of the quantum. We see that the escape probability is highest for the s-wave, so we set $l=0$. Then the expected time after which the trapped quantum will escape is
\be
\Delta t_{escape}=P_0^{-1}\Delta t_{osc}
\ee
The low energy quanta in the throat have $\omega'\sim 2\pi/\Delta t_{osc}$ \cite{lm4,lm6} so for our estimate we set
\be
\omega'={2\over R'_y}
\ee
We then find
\be
\alpha\equiv {\Delta t_{escape}\over \Delta t_{osc}}~=~{1\over (2\pi)^2} {R'^4_y\over {\bar Q}'_1{\bar Q}'_5}~=~{1\over (2\pi)^2}\, [{({\bar Q}'_1{\bar Q}'_5)^{1\over 4}\over a'}]^4
\ee
where $a'=({\bar Q}'_1{\bar Q}'_5)^{1/2}/R'_y$ is the radius obtained from $a$ after the dualities to the D1-D5 frame \cite{lm4}. In this frame the cap+throat region has the geometry of global $AdS_3\times S^3\times T^4$. The curvature radius of the $AdS_3$ and $S^3$ is $({\bar Q}'_1{\bar Q}'_5)^{1/4}$. The ratio 
\be
\beta\equiv {({\bar Q}'_1{\bar Q}'_5)^{1\over 4}\over a'}
\ee
gives the number of $AdS$ radii that we can go outwards from $r=0$ before reaching the `neck' region.\footnote{This can be seen from the metric for the profile (\ref{profile}) \cite{bal,mm,fuzz}.} Thus $\beta$ is a measure of the depth of the throat compared to its diameter.

While all lengths in the noncompact directions are scaled under the dualities, the ratio of such lengths is unchanged.
Thus in the NS1-P duality frame
\be
\beta = {({\bar Q}_1{\bar Q}_p)^{1\over 4}\over a}
\ee
We note that
\be
\alpha\sim \beta^4
\label{slow}
\ee
Thus when the throat becomes deep the quanta trapped in the throat become long lived excitations of the system.

For completeness let us also start from the other limit, where the coupling is weak and we have a thin long tube as in Fig.\ref{fig2m}(b). 
The radius of the ring described by (\ref{profile}) is $a$. The gravitational effect of NS1,P charges extends to distances $Q_1, Q_p$ from the ring. In the definitions (\ref{qqbar}) we put in the profile (\ref{profile}), and find 
\be
Q_1={\bar Q_1\over 2a}, ~~~~~~ Q_p={\bar Q_p\over 2a}
\label{qqqq}
\ee
If we take for the `thickness' of the ring the length scale $\sqrt{Q_1Q_p}$ then from (\ref{qqqq}) we find
\be
{(Q_1Q_p)^{1/2}\over a}={(\bar Q_1\bar Q_p)^{1/2}\over 2a^2}={\beta^2\over 2}
\ee
so we see again that the ring `thickness' becomes comparable to the ring radius when $\beta\sim 1$. For $\beta \ll 1$ we have a `thin ring' and for $\beta \gg 1$ we have a `deep throat'. 

Instead of using $\sqrt{Q_1Q_p}$ as a measure of the ring thickness we can say that the ring is thin when 
\be
a\gtrsim  Q_1,~~~a\gtrsim Q_p
\ee
This can be encoded in the requirement
\be
a\gtrsim {Q_1Q_p\over Q_1+Q_p}
\label{thin}
\ee
From the first equality in (\ref{periodf}) we get an expression for $\Delta t$, which we equate to the expression found in  (\ref{periodp}), to get 
\be
{1\over 2}L_z {Q_1+Q_p\over \sqrt{Q_1Q_p}}=\Delta t=\pi\alpha' M_T
\ee
where $L_z=2\pi a$ is the length of the ring and $M_T$ is its total mass. Using this  and (\ref{qqqq}) we can rewrite (\ref{thin}) as
\be 
\alpha' M_T\gtrsim (Q_1Q_p)^{1/2}={(\bar Q_1\bar Q_p)^{1/2}\over 2a}
\ee
Expressing the macroscopic parameters in terms of the microscopic charges and moduli\footnote{The expression for $a$ is  obtained by using the  profile (\ref{profile}) in (\ref{qbarp}) and use the expressions (\ref{integercharges}).}
\be
\bar Q_1={g^2\alpha'^3n_1\over V}, ~~~\bar Q_p={g^2 \alpha'^4n_p\over VR_y^2}, ~~~a=\sqrt{n_1n_p\alpha'}
\ee
we find that the ring is `thin' when
\be
\alpha' M_T\gtrsim {g^2\alpha'^{3}\over VR_y}
\label{con1}
\ee
This version of the criterion for ring thickness will be of use below.

\subsection{The phase transition in the microscopic picture}

We now turn to the microscopic description of the system.
Consider first the BPS bound state in the D1-D5 duality frame. Suppose we add a little bit of energy to take the system slightly above extremality.  From the work on near-extremal states \cite{callanmalda,maldastrom,lm4,review} we know that the energy will go to exciting vibrations that run up and down the components of the effective string
\be
D1-D5~+~\Delta E ~~\r~~D1-D5~+~P\bar P
\label{pr1}
\ee
where we call the excitations $P\bar P$ since they carry momentum charge in the positive and negative $S^1$ directions.
For the geometry made by starting with the NS1-P profile (\ref{profile}) we have no `fractionation'; i.e. the effective string formed in the D1-D5 bound state has $n_1n_5$ `singly wound' circles \cite{lm4}. Thus the minimum energy needed to excite the system is the energy of one left and one right mover on the effective string
\be
\Delta E^{D1D5}_{P\bar P}={1\over R'_y}+{1\over R'_y}={2\over R'_y}
\label{energy1}
\ee
The charges D1-D5-P can be permuted into each other, so we can map D1-D5 to P-D1, and then the dual of (\ref{pr1}) is
\be
P-D1~+~\Delta E ~~\r~~P-D1~+~D5\overline {D5}
\ee
A further S duality brings the system to the NS1-P system that we are studying, and then we get
\be
P-NS1~+~\Delta E ~~\r~~P-NS1~+~NS5\overline {NS5}
\label{pr2}
\ee
This may look strange, since it says that if we excite an oscillating string the energy of excitations goes to creating pairs of NS5 branes; we are more used to the fact that energy added to a string just creates more oscillations of the string. Dualizing 
(\ref{energy1}) gives for the excitation (\ref{pr2}) the minimum energy threshold
\be
\Delta E^{NS1P}_{NS5\overline{NS5}}=2 m_{NS5}=2{VR_y\over g^2\alpha'^3}
\ee
Thus at small $g$ these excitations are indeed heavy and should not occur. For comparison, we  find the minimum energy required to excite {\it oscillations} on the NS1-P system.  For small $g$ we use the spectrum of the free string which gives
\be
M^2=( {R_yn_1\over \alpha'}    +{ n_p\over R_y})^2+{4\over \alpha'}  N_L=( {R_yn_1\over \alpha'}    -{ n_p\over R_y})^2+{4\over \alpha'}  N_R
\label{massformula}
\ee
The lowest excitation is given by $\delta N_L=\delta N_R=1$. This gives
\be
\Delta E^{NS1P}_{oscillations}=\Delta M={2\over \alpha'}{1\over M_T}
\ee
where $M_T$ is the total mass of the BPS NS1-P state.  

We now observe that  oscillations on the NS1-P system are lighter than NS5 excitations only when
\be
 {1\over \alpha' M_T}\lesssim {VR_y\over g^2\alpha'^3}
 \label{con2}
\ee
Thus for very small $g$ the lightest excitation on the NS1-P system  is  an oscillation of the string. But   above a certain $g$ the $NS5\overline{NS5}$ pairs are lighter  and so will be the preferred  excitation when we add energy to the system.

\subsection{Comparing the gravity and microscopic pictures}

We now observe that the conditions (\ref{con1}) and (\ref{con2}) are the {\it same}. Thus we see that when the ring is thin then in the corresponding microscopic picture we have `2-charge excitations'; i.e. the third charge NS5 is not excited and the string giving the NS1-P state  just gets additional excitations which may be interpreted as pairs of NS1 and P charges. But when we increase the coupling beyond the point where the ring becomes `thick' and the geometry is better described as a throat, then the dual CFT has `3-charge excitations' which are pairs of NS5 branes. When $g$ is small and the ring is thin then the oscillations of the supertube are long lived because they couple only weakly to the radiation modes of the gravity field. When the tube becomes very {\it thick} then the oscillation modes are again long lived -- we get $\beta \gg 1$ and by (\ref{slow}) this implies a very slow leakage of energy to infinity.  

Thus we see that the modes at small and large coupling should not be seen as the `same' modes; rather the `2-charge modes' at weak coupling disappear at larger $g$ because of coupling to the radiation field, and at still larger $g$ the `3-charge modes' appear. For these latter modes one might say that the gravitational field of the system has `trapped' the excitations of the metric from the region ${\bar r}
\lesssim (\bar Q_1)^{1/2}, (\bar Q_p)^{1/2}$, so that these modes have in some sense been extracted from the radiation field. 

\section{A conjecture on identifying bound states for the 3-charge extremal system}\label{acon}
\setcounter{equation}{0}

Consider a D0 brane placed near a D4 brane. The force between the branes vanishes. But now give the D0 a small velocity in the space transverse to the D4. The force between the branes goes as $\sim v^2$, and the motion of the D0 can be described as a geodesic on the moduli space of its static configurations \cite{dkps}. This moduli space would be flat if we took a D0-D0 system (which is 1/2 BPS) but for the D0-D4 case (which is 1/4 BPS) the metric is a nonflat hyperkahler metric.

We can look at more complicated systems, for example 3-charge black holes in 4+1 spacetime. Now the system is 1/8 BPS. The positions of the black holes give coordinates on moduli space, and the metric on moduli space was computed in \cite{kami}. If we set to zero one of the three charges then we get a 1/4 BPS system.

It is easy to distinguish `motion on moduli space' from the kinds of oscillatory behavior that we have encountered in the dynamics of supertubes. As mentioned in the introduction, when we have motion on moduli space we take the limit of the velocity going to zero, and  over a long time $\Delta t$ the system configuration changes by order unity. Using $\Delta x$ as a general symbol for the change in the configuration\footnote{For example $x$ could be the separation of two black hole centers.} we have for `drift on moduli space'
\be
v\sim \epsilon, ~~~\Delta t\sim {1\over \epsilon}, ~~~\Delta x\sim 1, ~~~~~~~(\epsilon\r 0)
\label{driftq}
\ee
On the other hand for the periodic behavior  that we have found for  both the weak coupling and  strong coupling dynamics of bound states, we have
\be
v\sim \epsilon, ~~~\Delta t\sim 1, ~~~\Delta x\sim \epsilon ~~~~~~~(\epsilon\r 0)
\label{quasiq}
\ee
Note that for the motion (\ref{driftq}) the energy lost to radiation during the motion vanishes as $\epsilon\r 0$, so the dynamics (\ref{driftq}) is unlike any of the cases that we have discussed for the bound state.

While the moduli space metric in \cite{kami} was found for spherically symmetric black holes (`naive geometries' in the language of \cite{lm4}) we expect that a similar `drifting' motion would occur even if we took two `actual' geometries of the 2-charge system and gave them a small relative velocity with respect to each other.\footnote{The motion of the centers of the two states could be accompanied by a slow change in the internal configurations of the states.} Thus such unbound systems would have a dynamical mode {\it not} present for the bound states. 

For the bound state 3-charge geometries that have been constructed \cite{3charge} the structure is very similar to the structure of 2-charge geometries. 
It is therefore reasonable to conjecture that 3-charge geometries will have a similar behavior: Unbound systems will have `drift' modes like (\ref{driftq}) while bound systems will have no such modes. If true, this conjecture could be very useful for the following reason. It is known how to write down the class of {\it all} 3-charge supersymmetric geometries \cite{gmr,benawarner,gauntlett}. But we do not know which of these are bound states. On the other hand the microstates of the 3-charge black hole \cite{sv} are bound states of three charges. If we can select the bound state geometries  from the unbound ones by some criterion then we would have a path to understanding all the microstates of the 3-charge black hole. This is important because the 3-charge hole has a classical horizon and our results on this hole should extend to   all holes.

To summarize, it seems a reasonable conjecture that out of the class of all supersymmetric 3-charge geometries the bound states are those that have no `drift' modes (\ref{driftq}). It would be interesting to look for `drift' modes for the 3-charge geometries constructed recently in \cite{gimon,bw2}; here the CFT dual is not known so we do not know a priori if the configuration is a bound state. The same applies for geometries made by adding KK-monopole charge to BPS systems carrying  a smaller number of charges \cite{kk}. It would also be useful to extend these considerations to the suggested construction of 3-charge supertubes and their geometries \cite{bekr,bena}.

In the introduction we have also asked the question: Can the bound state break up into two or more unbound states under a small perturbation? Since the bound state is only threshold bound, such a breakup is allowed on energetic grounds. But for the 2-charge system we see  that bound states are not `close' to unbound states. The bound states are described by a simple closed curve traced out by the locations $\vec x=\vec F(v)$ for $0\le v < L_T$. A superposition of {\it two} such bound states has {\it two} such simple closed curves. The curve can break up into two curves if it self-intersects, but in a generic state the curve is not  self-intersecting. If we add a little energy to a 2-charge bound state then we have seen that the configuration does not `drift' through the space of bound states, so the curve will not drift  to a curve with a self-intersection and then split. Thus we expect that generic bound states are stable to small perturbations; energy added to them causes small oscillations for a while and the energy is eventually lost to infinity as radiation.

\section{Discussion}

Let us summarize our arguments and conjectures. If we have two BPS objects with 1/4 susy then they feel no force at rest, but their low energy dynamics is a slow relative motion described by geodesics on a moduli space. If we look at just one 1/4 BPS bound state then it has a large degeneracy, which in the classical limit  manifests itself as a continuous family 
of time-independent solutions. If we add a small energy to the BPS bound state, then what is the evolution of the system?

Based on the behavior of unbound objects one might think that there will again be a `drift' over the family of configurations, described by some metric on the moduli space of configurations. But we have argued that this is not what we should expect. We first looked at the 1/4 BPS configurations at zero coupling, where we get `supertubes' described by a DBI action. We saw that the best way to get the dynamics of such 1/4 BPS objects is to use the NS1-P picture, which is a `multiwound string carrying  a traveling wave'. For this zero coupling limit we found that instead of a `drift' over configurations' we get oscillatory behavior. These oscillations are not described by a collection of  simple harmonic oscillators. Rather they are like the motion of a charged particle in a magnetic field where each term in the equation of motion has at least one time derivative, and there is a continuous family of equilibrium configurations. We found a simple expression (\ref{periodp}) for the period of  oscillations with arbitrary amplitude, which reduced to the period found in \cite{pm} for the case of  small oscillations of the round supertube.

If we increase the coupling a little then we get a gravity description of the supertube, but with gravitational field of the tube extending only to distances small compared to the circumference of the tube. Thus we get a `thin' long tube. Zooming in to a point of the tube we see an essentially straight segment, and we studied the perturbations to this geometry. We found excitations that agree in frequency with those found from the zero coupling analysis.

We noted that the part of the excitation that leaks out to spatial infinity will have the form of a traveling wave. As we increase the coupling the amplitude of the wave reaching this region becomes larger.  Thus there will be an energy flux leaking out to infinity, and the excitation will not remain concentrated near the supertube. But as we increase the coupling still further we find that the geometry develops
a deep `throat' and we get a new kind of long lived excitation: Supergravity modes can be trapped in this throat for long times, only slowly leaking their energy to infinity.

We argued that the different kinds of excitations found at weak and strong coupling reflect the phase transition that had been noted earlier from the study of black holes \cite{emission,review}. At weak coupling the excitations on such a system creates pairs of the charges already
present in the BPS state. But at larger coupling the excitation energy goes to creating pairs of a {\it third} kind of charge.
The value of $g$ where this transition occurs was found to have the same dependence on $V, R, n_i$ as the value of $g$ where the supertube stops being `thin';  i.e. where the gravitational effect of the tube starts extending to distances comparable to the radius of the tube.

We have noted that bound states do not exhibit a `drift' over a moduli space of configurations, while unbound states do. If 3-charge systems behave qualitatively in the same way as 2-charge ones then this fact can be used to distinguish bound states from unbound ones for the class of 1/8 BPS states; such  bound states would give microstates of the 3-charge extremal hole.

\section*{Acknowledgements}

 This work is
supported in part by DOE grant DE-FG02-91ER-40690. S.G. was supported by an I.N.F.N. fellowship.
We would like to thank Camillo Imbimbo and Ashish Saxena for helpful discussions.
  
\appendix

\section{Analysis of the  $\mathcal{A}^{+}$ equations}

\renewcommand{\theequation}{A.\arabic{equation}}
\setcounter{equation}{0}

In this appendix we look at the field equations (\ref{eqa})  for the field $\mathcal{A}^{+}$. We have found the expected time-independent solutions in section (\ref{secaplus}). We will now consider a general ansatz for the solution and argue that there are no time {\it dependent} solutions for this field, if we demand consistency with the long wavelength limit  (\ref{straight}).

Let us write 
 \be
\mathcal{A}^{+}_v=H^{-1}\,a^{+}_v\,,\quad \mathcal{A}^{+}_t=a^{+}_t\,,\quad  \mathcal{A}^{+}_z=a^{+}_z\,,\quad 
 \mathcal{A}^{+}_i=a^{+}_i
\ee
Since we have spherical symmetry in the space spanned by the coordinates $i$ all fields will be functions only of the radial coordinate $r$ in this space; further, the $\mathcal{A}^+_i$ can have only the component $\mathcal{A}^+_r$. Putting this ansatz into (\ref{eqa}) we obtain the coupled system of equations (we list the equations in the order $\lambda=t,v,z,i$)
\bea
\triangle\, a^{+}_t+\partial_t^2 a^{+}_v -\partial_t\partial_i a^{+}_i 
+\partial_z(\partial_z a^{+}_t-\partial_t a^{+}_z)
+A\,\partial_t(\partial_z a^{+}_t -\partial_t a^{+}_z)=0
\label{eqt+}
\eea

\bea
&&\triangle\, a^{+}_v+\partial_z^2 a^{+}_v-[(H\Kt -A^2)\,\partial_t^2 a^{+}_v - 2A\,
\partial_t\partial_z a^{+}_v]\nonumber\\
&&+[\partial_i (H\Kt -A^2)\partial_i a^{+}_t -2 \partial_i A\,\partial_i a^{+}_z]-
[\partial_i (H\Kt -A^2)\partial_t a^{+}_i -2 \partial_i A\,\partial_z a^{+}_i]=0\nonumber\\
\label{eqv+}
\eea

\bea
&&\triangle\, a^{+}_z+\partial_t\partial_z a^{+}_v -\partial_z\partial_i a^{+}_i +  (H\Kt -A^2)\,
\partial_t(\partial_z a^{+}_t - \partial_t a^{+}_z)\nonumber\\
&& -A\,\partial_z(\partial_z a^{+}_t -\partial_t a^{+}_z)=0
\label{eqz+}
\eea

\bea
&&\partial_t\partial_i a^{+}_v-\partial_z\partial_i a^{+}_z+\partial_z^2 a^{+}_i -\partial_j(\partial_i a^{+}_j -\partial_j a^{+}_i)+\partial_i A\,
(\partial_z a^{+}_t -\partial_t a^{+}_z)\nonumber\\
&&+
[(H\Kt -A^2)\,\partial_t\partial_i a^{+}_t - A\,
\partial_z\partial_i a^{+}_t-A\,\partial_t\partial_i a^{+}_z]\nonumber\\
&&-[(H\Kt -A^2)\partial_t^2 a^{+}_i - 2 A
\partial_t\partial_z a^{+}_i]
=0
\label{eqi+}
\eea

We look for a gauge field $\mathcal{A}^{+}$ having the same $z$ and $t$ dependence as $\mathcal{A}^{-}$. We write
\bea
&&
a^{+}_v=e^{i k\, z- i\omega\, t}\,f_v(r)\,,\quad a^{+}_t=e^{i k\, z- i\omega\, t}\,f_t(r)\nonumber\\
&&a^{+}_z=e^{i k\, z-i\omega\, t}\,f_z(r)\,,\quad a^{+}_r=i e^{i k\, z- i\omega\, t}\,\partial_r \Lambda
\label{ansatz+}
\eea
where the notation we have used for $a^+_r$ will be helpful in what follows.

Consider first equations (\ref{eqt+}) and (\ref{eqz+}):
\bea
&&\triangle\, f_t -\omega^2\,f_v - \omega\,\triangle \Lambda -(k\,f_t+\omega\,f_z)\,(k-\omega\,A)=0\nonumber\\
&&\triangle\, f_z +\omega k\,f_v + k\,\triangle\, \Lambda +(k\,f_t+\omega\,f_z)\,[k\,A +\omega\,(H\Kt -A^2)]=0
\eea
Taking $k$ times the first equation plus $\omega$ times the second we get
\be
\triangle\, (k\,f_t+\omega\,f_z)- [k^2 - 2\omega k\,A - \omega^2 (H\Kt -A^2) ]\,(k\,f_t+\omega\,f_z)=0
\ee
We note that the coefficient $k^2 - 2\omega k\,A - \omega^2 (H\Kt -A^2)$ has the form $c+d/r$. If $d\ne 0$ then by an argument similar to that leading to (\ref{omega}) we find that there is no solution with the required behavior at $r=0$. Setting  $d=0$ tells us that
we need to have the same relation between $\omega$ and $k$ that we have seen before (\ref{omega})
\be
\omega=-k\,{2\sqrt{Q_1 Q_p}\over Q_1+Q_p}
\label{omegap}
\ee
Using (\ref{omegap}) one finds
\be
k^2 - 2\omega k\,A - \omega^2 (H\Kt -A^2)=k^2\,\Bigl({Q_1-Q_p\over Q_1+Q_p}\Bigr)^2~\equiv ~{\tilde k}^2
\ee
and thus
\be
\triangle\, (k\,f_t+\omega\,f_z)- {\tilde k}^2\,(k\,f_t+\omega\,f_z)=0
\ee
The solution of the above equation that converges at infinity is
\be
k\,f_t+\omega\,f_z={\tilde c}\,{e^{-|{\tilde k}|r}\over r}
\label{kf1}
\ee 

Let us now look at equation (\ref{eqv+}). Using (\ref{omegap}) and (\ref{kf1}) we find
\bea
&&[(H\Kt -A^2)\,\partial_t^2 a^{+}_v - 2A\,\partial_t\partial_z a^{+}_v]=-\omega^2\,f_v\nonumber\\
&&[\partial_i (H\Kt -A^2)\partial_i a^{+}_t -2 \partial_i A\,\partial_i a^{+}_z]=-{Q_1+Q_p\over r^2}\,{{\tilde c}\over k}\,\partial_r\Bigl({e^{-|{\tilde k}|r}\over r}\Bigr)\nonumber\\
&&[\partial_i (H\Kt -A^2)\partial_t a^{+}_i - 2 \partial_i A\,\partial_z a^{+}_i]=0
\eea 
Then equation (\ref{eqv+}) becomes
\be
\triangle f_v -{\tilde k}^2\,f_v - {Q_1+Q_p\over r^2}\,{{\tilde c}\over k}\,\partial_r\Bigl({e^{-|{\tilde k}|r}\over r}\Bigr)=0
\label{fv1}
\ee
We find that unless 
\be
{\tilde c}=0
\label{tc1}
\ee
$f_v$ will be too singular to agree with  (\ref{straight}) at small $r$. The vanishing of $\t c$ also makes (\ref{kf1}) agree with
(\ref{straight}) at small $r$.

From (\ref{fv1}) and using the short distance limit implied by (\ref{straight}) we get
\be
f_v = ({\tilde \alpha}-{\tilde \beta})\,Q_1\,{e^{-|{\tilde k}|r}\over r}
\label{fv2}
\ee
Consider now the last equation (\ref{eqi+}) (for $i=r$, the only non-trivial component). The fact that
${\tilde c}=0$ implies
\be
\partial_z a^{+}_t -\partial_t a^{+}_z=0
\ee
Moreover one has
\bea
&&[(H\Kt -A^2)\,\partial_t\partial_r a^{+}_t - A\,
\partial_z\partial_r a^{+}_t-A\,\partial_t\partial_r a^{+}_z]=-i \omega\,\partial_r f_t\nonumber\\
&&[(H\Kt -A^2)\partial_t^2 a^{+}_r - 2 A \partial_t\partial_z a^{+}_r]=-i \omega^2 \partial_r \Lambda
\eea
Equation (\ref{eqi+}) then gives
\be
\partial_r f_v + \Bigl(1-{k^2\over \omega^2}\Bigr)\,\partial_r(f_t-\omega\,\Lambda)=0
\label{rbis}
\ee
Equation (\ref{eqt+}) also simplifies to
\be
\triangle\, (f_t-\omega\,\Lambda)-\omega^2\,f_v=0
\label{tbis}
\ee
We see that (\ref{rbis}) implies (\ref{tbis}).  But at this stage we would like to compare the solution we have found with the  limits required by (\ref{straight}). 
First consider the case $Q_1=Q_p$. Then  $\omega=k$ and (\ref{rbis}) is not compatible with  (\ref{fv2}). Now consider $Q_1\not = Q_p$. Eq.~(\ref{rbis}) implies  
\be
f_t-\omega\,\Lambda = {\omega^2\over k^2-\omega^2}\,f_v + {\rm const.}
\ee
This is again incompatible with the limit (\ref{straight}), according to which $f_t-\omega\,\Lambda$ should {\it vanish} for small $r$.
We conclude that there are no time-dependent solutions for $\mathcal{A}^{+}$ consistent
with the limit (\ref{straight}). 

\section{Asymptotic behavior of the perturbation} \label{asym}
\setcounter{equation}{0}

\renewcommand{\theequation}{B.\arabic{equation}}
\setcounter{equation}{0}

In this appendix we will study the behavior of perturbations on the 2-charge geometries near spatial infinity. We wish to see how fields fall off with $\bar r$, and in particular to check that they carry energy flux off to infinity. We will first look at a scalar field to get an idea of the behaviors involved, and then address the 1-form gauge field that is actually excited in our problem.

\subsection{The scalar perturbation}

Consider first the case in which the perturbation is represented by a minimally coupled scalar
$\Psi$. We take the metric (\ref{fp}) and look at its large $\bar r$ limit. 
We denote by $\bar r$, $\theta$, $\phi$, $\psi$ the spherical coordinates 
in the 4 noncompact dimensions ${\bar x}_i$ ($i=1,\ldots,4$). 
The 5D Einstein metric is
\be
ds^2=-h^{-4/3}\,dt^2+h^{2/3}\,(d{\bar r}^2+{\bar r}^2\, d\theta^2+{\bar r}^2\,\sin^2\theta\,d\phi^2 +
{\bar r}^2\,\cos^2\theta\,d\psi^2)
\ee
with
\be
h=\sqrt{\Bigl(1+{\Qbar\over {\bar r}^2}\Bigr)\Bigl(1+{\QPbar\over {\bar r}^2}\Bigr)}
\ee 
We set momentum along the $S^1$ to zero and expand in angular harmonics
\be
\Psi=e^{-i\omega\,t}\,\mathcal{R}({\bar r})\,Y^{(l)}(\theta,\phi,\psi)
\ee
where $Y^{(l)}$ the $l$-th scalar spherical harmonic. 
The wave equation for $\Psi$
\be
\Box \Psi=0
\ee
implies
\be
{1\over {\bar r}^3}\partial_{\bar r}({\bar r}^3 \partial_{\bar r} \mathcal{R})+\omega^2\,\Bigl(1+{\Qbar+\QPbar\over {\bar r}^2}+{\Qbar \QPbar\over {\bar r}^4}
\Bigr)\mathcal{R} -{l(l+2)\over {\bar r}^2}\,\mathcal{R}=0
\label{equation7}
\ee
We can understand the behavior of $\mathcal{R}$ at large ${\bar r}$ as follows.  
If we define
\be
\mathcal{R}={\tilde \mathcal{R}\over {\bar r}^{3/2}}
\ee
the equation for $\tilde \mathcal{R}$ is
\be
\partial_{\bar r}^2 {\tilde \mathcal{R}}+\omega^2\,{\tilde \mathcal{R}}+{(\Qbar+\QPbar)\omega^2-l(l+2)-3/4\over {\bar r}^2}\,{\tilde \mathcal{R}}+
{\Qbar \QPbar\over {\bar r}^4}\omega^2\,{\tilde \mathcal{R}}=0
\label{eqR}
\ee
At leading order in $1/{\bar r}$ the terms proportional to $1/{\bar r}^2$ and $1/{\bar r}^4$ can be neglected and we have
the solution
\be
{\tilde \mathcal{R}}=r_{+}\,e^{i \omega\,{\bar r}}+r_-\,e^{-i \omega\,{\bar r}}\,\,~~\Rightarrow\,\,~~
\mathcal{R}={r_{+}\,e^{i \omega\,{\bar r}}+r_-\,e^{-i \omega\,{\bar r}}\over {\bar r}^{3/2}}
\ee
which corresponds to traveling waves carrying a nonzero flux. When
the terms of higher order in $1/{\bar r}$ are included, (\ref{eqR}) can be recursively
solved as a formal power series in $1/{\bar r}$:
\be
\mathcal{R}~\approx~ r_{+}\,{e^{i \omega\,{\bar r}}\over {\bar r}^{3/2}}\,\Bigl(1+
\sum_{n=1}^\infty{r_+^{(n)}\over {\bar r}^n}\Bigr)+
r_{-}\,{e^{-i \omega\,{\bar r}}\over {\bar r}^{3/2}}\,\Bigl(1+\sum_{n=1}^\infty
{r_-^{(n)}\over {\bar r}^n}\Bigr)
\label{expR}
\ee
The coefficients $r_\pm^{(n)}$ in this expansion are determined by the
recursion relation
\be
r_\pm^{(n)} = \mp{i\over 2\omega}\,\Bigl[r_\pm^{(n-1)}\,\Bigl(n-1+
{(\Qbar+\QPbar)\omega^2-l(l+2)-3/4\over n}\Bigr)+r_\pm^{(n-3)}\,
{\Qbar\QPbar\,\omega^2\over n}\Bigr]
\ee 
The $r_\pm^{(n)}$ are finite for any value of $\omega$ and any 
 $n$. However, since at large $n$ one has
\be
{r_\pm^{(n)}\over r_\pm^{(n-1)}}\approx\mp{i(n-1)\over 2\omega}
\ee
the series (\ref{expR}) has zero radius of convergence. Equations like (\ref{equation7}) lead instead to asymptotic series in $1/r$ \cite{book},  and  we expect that the above expansion  is to be interpreted as an 
asymptotic series, which accurately describes the behavior of 
$\mathcal{R}$ at sufficiently large ${\bar r}$.
From (\ref{expR}) we can still conclude that the perturbation $\mathcal{R}$ 
radiates a finite amount of flux at infinity. 
Note that, in order to avoid logarithms in the expansion (\ref{expR}), it is
crucial that the next to leading corrections to the equation (\ref{eqR})
are of order $1/{\bar r}^2$.

\subsection{The vector perturbation}

We find a similar situation for the case in which the perturbation is 
represented by a vector on the six dimensional space $\mathbb{R}^{(5,1)}\times
S^1$. As we showed in section (\ref{grav}) the perturbation on the 2-charge system is a vector field with wave equation
\be
\nabla_\mu (e^{-2\Phi}\,F^{\pm\,\mu\lambda})\pm {1\over 2}e^{-2\Phi}\,H^{\mu\nu\lambda}\,F^\pm_{\mu\nu}=0
\label{eqasym}
\ee
The gauge fields $\mathcal{A}^+_\mu$ and $\mathcal{A}^-_\mu$ represent respectively BPS 
and non-BPS perturbations. Since we are interested in time-dependent, non-BPS, perturbations, we will only look at the equation for $\mathcal{A}^-_\mu$ in this section. 
For the metric, dilaton and B-field appearing in
(\ref{eqasym}) we will take the large ${\bar r}$ limits
\bea
ds^2&=&{\bar H}^{-1}\,[-dt^2+dy^2+{\bar K}\,(dt-dy)^2]+d{\bar r}^2+{\bar r}^2\,d\theta^2+{\bar r}^2\,\sin^2\theta\,d\phi^2+{\bar r}^2\,\cos^2\theta\,d\psi^2\nonumber\\
B&=&-({\bar H}^{-1}-1)\,dt\wedge dy\,,\quad e^{2\Phi}={\bar H}^{-1}
\label{naive}
\eea
with
\be
{\bar H}=1+{\Qbar\over {\bar r}^2}\,,\quad {\bar K}={\QPbar\over {\bar r}^2}
\ee
The spherical symmetry of the background (\ref{naive}) allows us to expand
the vector field components into spherical harmonics: Denoting by
$Y^{(l)}$ and $Y^{(l)}_\alpha$ the scalar and vector spherical harmonics on 
$S^3$, we can write
\be
\mathcal{A}^-_I=e^{-i\omega\,t}\,\mathcal{R}_I({\bar r})\,Y^{(l)}(\theta,\phi,\psi)\,,\quad
\mathcal{A}^-_\alpha=e^{-i\omega\,t}\,[\mathcal{R}_s({\bar r})\,\partial_\alpha 
Y^{(l)}(\theta,\phi,\psi)+\mathcal{R}_v({\bar r})\,Y^{(l)}_\alpha(\theta,\phi,\psi)]
\ee
with $I=t,y,{\bar r}$ and $\alpha=\theta,\phi,\psi$. We will need the following
spherical harmonic identities 
\bea
\Box' Y^{(l)}=-l(l+2)\,Y^{(l)}\,,\quad 
\Box' Y^{(l)}_\alpha=(2-(l+1)^2)\,Y^{(l)}_\alpha\equiv -c(l)\,Y^{(l)}_\alpha\,,\quad \nabla'^\alpha
Y^{(l)}_\alpha=0 
\eea
where ``primed'' quantities refer to the metric on an $S^3$ of unit radius. (We use a notation in which
$l=0,1,\ldots$ for the scalar harmonics and $l=1,2,\ldots$ for the vector harmonics). The
components with $\lambda=\alpha$ in (\ref{eqasym}) give 
\bea
&&{1\over {\bar r}}\,\partial_I\Bigl[{\bar r}\,g^{IJ}\,\partial_J (e^{-i\omega\,t}\,\mathcal{R}_v)\Bigr]-
{c(l)+2\over {\bar r}^2}\,e^{-i\omega\,t}\,\mathcal{R}_v=0\nonumber\\
&&{1\over {\bar r}}\,\partial_I\Bigl[{\bar r}\,g^{IJ}\,
\Bigl(e^{-i\omega\,t}\,\mathcal{R}_J-\partial_J (e^{-i\omega\,t}\,\mathcal{R}_s)\Bigr)
\Bigr]=0
\label{eqalpha}
\eea
and the components with $\lambda=I$ give
\bea
\label{eqI}
&&\!\!\!\!\!\!\!\!\!\!\!{1\over {\bar r}^3}\partial_K\Bigl[{\bar r}^3\,g^{K L}\,g^{IJ}
\Bigl(\partial_L(e^{-i\omega\,t}\,\mathcal{R}_J)-\partial_J(e^{-i\omega\,t}\,\mathcal{R}_L)\Bigr)\Bigr]\\
&&\!\!\!\!\!\!\!\!\!\!\!-{g^{IJ}\,l(l+2)\over {\bar r}^2}
\Bigl(e^{-i\omega\,t}\,\mathcal{R}_J-\partial_J (e^{-i\omega\,t}\,\mathcal{R}_s)\Bigr)
- \epsilon^{IJK}{\Qbar\over {\bar r}^3} \Bigl(\partial_J(e^{-i\omega\,t}\,\mathcal{R}_K)-
\partial_K(e^{-i\omega\,t}\,\mathcal{R}_J)\Bigr)=0\nonumber
\eea
with $\epsilon^{{\bar r}ty}=1$. As expected from group theory considerations, the 
component $\mathcal{R}_v$ decouples from all others, while $\mathcal{R}_s$ and $\mathcal{R}_I$
satisfy a coupled system of differential equations. We want to show that, in spite of these mixings, $\mathcal{R}_v$, $\mathcal{R}_s$ and $\mathcal{R}_I$ admit an $1/{\bar r}$
expansion analogous to (\ref{expR}). Putting in the explicit value of $g_{IJ}$ in  (\ref{eqalpha}) and (\ref{eqI}) and using the gauge 
\be
\mathcal{A}^-_t=0
\ee
we obtain the following system of equations
\bea
{1\over {\bar r}}\partial_{\bar r}({\bar r}\partial_{\bar r} \mathcal{R}_v)+\omega^2\Bigl(1+{\Qbar+\QPbar\over {\bar r}^2}+{\Qbar \QPbar\over {\bar r}^4}\Bigr)\,\mathcal{R}_v-{c(l)+2\over {\bar r}^2}\,\mathcal{R}_v=0
\label{eqrv}
\eea
\bea
{1\over {\bar r}}\partial_{\bar r}({\bar r}\partial_{\bar r} \mathcal{R}_s)
+\omega^2\Bigl(1+{\Qbar+\QPbar\over {\bar r}^2}+{\Qbar \QPbar\over {\bar r}^4}\Bigr)\mathcal{R}_s-{1\over {\bar r}}\partial_{\bar r}({\bar r} \mathcal{R}_{\bar r})-i\omega{\QPbar\over {\bar r}^2}
\Bigl(1+{\Qbar\over {\bar r}^2}\Bigr)\mathcal{R}_y=0\nonumber\\
\label{eqrs}
\eea
\bea
&&{l(l+2)\over {\bar r}^2}\,(\partial_{\bar r} \mathcal{R}_s -\mathcal{R}_{\bar r})
+\omega^2\Bigl(1+{\Qbar+\QPbar\over {\bar r}^2}+{\Qbar \QPbar\over {\bar r}^4}\Bigr)\,
\mathcal{R}_{\bar r}\nonumber\\
&&\quad -i\omega\,{\QPbar\over {\bar r}^2}
\Bigl(1+{\Qbar\over {\bar r}^2}\Bigr)\,\partial_{\bar r} \mathcal{R}_y+2i\omega\,{\Qbar\over {\bar r}^3}\,\mathcal{R}_y=0\label{eqrr}
\eea
\bea
&&{1\over {\bar r}^3}\partial_{\bar r} ({\bar r}^3 \mathcal{R}_{\bar r})-{l(l+2)\over {\bar r}^2}\,\mathcal{R}_s+i
\omega\,{\QPbar\over {\bar r}^2}\Bigl(1+{\Qbar\over {\bar r}^2}\Bigr)\,\mathcal{R}_y\nonumber\\
&&\quad-{2\over {\bar r}^3}\,
\Bigl(1+{\Qbar\over {\bar r}^2}\Bigr)^{-1}(\Qbar+\QPbar)\,\Bigl(\mathcal{R}_{\bar r}-{i\over \omega}\,
\partial_{\bar r} \mathcal{R}_y\Bigr)=0
\label{eqrt}
\eea
\bea
&&{1\over {\bar r}^3}\partial_{\bar r} ({\bar r}^3\partial_{\bar r} \mathcal{R}_y)-{l(l+2)\over {\bar r}^2}\,\mathcal{R}_y+\omega^2\Bigl(1+{\Qbar+\QPbar\over {\bar r}^2}+{\Qbar \QPbar\over {\bar r}^4}\Bigr)\,
\mathcal{R}_y\nonumber\\
&&\quad-{2\over {\bar r}^3}\,
\Bigl(1+{\Qbar\over {\bar r}^2}\Bigr)^{-1}(\Qbar-\QPbar)\,\Bigl(i\omega\,\mathcal{R}_{\bar r}+
\partial_{\bar r} \mathcal{R}_y\Bigr)=0
\label{eqry}
\eea
(Eq.~(\ref{eqrr}) is the $I={\bar r}$ component of (\ref{eqI}); eqs.~(\ref{eqrt}) and (\ref{eqry}) are linear combinations of the $I=t,y$ components of (\ref{eqI}).)

Eq.~(\ref{eqrv}) for $\mathcal{R}_v$ is analogous to eq.~(\ref{eqR}): it thus admits
an analogous asymptotic expansion, of the form 
\be
\mathcal{R}_v\approx r_{v,+}\,
{e^{i \omega\,{\bar r}}\over {\bar r}^{1/2}}\,\Bigl(1+\sum_{n=1}^\infty
{r_{v,+}^{(n)}\over {\bar r}^n}
\Bigr)+
r_{v,-}\,{e^{-i \omega\,{\bar r}}\over {\bar r}^{1/2}}\,\Bigl(
1+\sum_{n=1}^\infty{r_{v,-}^{(n)}\over {\bar r}^n}\Bigr)
\ee
When expressed in local orthonormal coordinates, the contribution of $\mathcal{R}_v$ to
$\mathcal{A}^-$ is of the type
\be
\mathcal{A}^-_{\hat\alpha}\sim {e^{\pm i \omega\,{\bar r}}\over {\bar r}^{3/2}}(1+O({\bar r}^{-1}))
\ee
and thus it again gives rise to a wave carrying finite flux at infinity.

The remaining eqs.~(\ref{eqrs}-\ref{eqry}) are four relations for the three
unknowns $\mathcal{R}_s$, $\mathcal{R}_{\bar r}$ and $\mathcal{R}_y$: this is so because we have used gauge invariance to eliminate one unknown, $\mathcal{R}_t$. It then must be that only three 
of the four eqs.~(\ref{eqrs}-\ref{eqry}) are linearly independent, and indeed 
one can check that eq.~(\ref{eqrt}), for example, follows from (\ref{eqrs}) and (\ref{eqrr}). 
We are thus left to solve the coupled system of equations (\ref{eqrs}), (\ref{eqrr}) and (\ref{eqry}).
We can do this by using the following strategy: solve eq.~(\ref{eqrr}) for $\mathcal{R}_{\bar r}$ and substitute into
(\ref{eqrs}) and (\ref{eqry}), which can then be solved for $\mathcal{R}_s$ and $\mathcal{R}_y$, iteratively in $1/{\bar r}$. 
To make the behavior at large ${\bar r}$ more transparent we also write $\mathcal{R}_s$ and $\mathcal{R}_y$ as
\be
\mathcal{R}_s={{\tilde \mathcal{R}}_s\over {\bar r}^{1/2}}\,,\quad \mathcal{R}_y={{\tilde \mathcal{R}}_y\over {\bar r}^{3/2}}
\ee  
We find
\be
\mathcal{R}_{\bar r} = -{l(l+2)\over \omega^2}\,{\partial_{\bar r} {\tilde \mathcal{R}}_s\over {\bar r}^{5/2}}+{1\over {\bar r}^{7/2}}\,
\Bigl({l(l+2)\over 2\omega^2}\,{\tilde \mathcal{R}}_s+i{\QPbar\over \omega}\,\partial_{\bar r} {\tilde \mathcal{R}}_y\Bigr)+
O({\bar r}^{-9/2})
\label{rr}
\ee
and
\bea
&&\partial_{\bar r}^2 {\tilde \mathcal{R}}_s + \omega^2\,{\tilde \mathcal{R}}_s+{\omega^2(\Qbar+\QPbar)+1/4\over {\bar r}^2}\,{\tilde \mathcal{R}}_s + 
{l(l+2)\over \omega^2\,{\bar r}^2}\,\partial_{\bar r}^2 {\tilde \mathcal{R}}_s + O({\bar r}^{-3})=0\nonumber\\
&&\partial_{\bar r}^2 {\tilde \mathcal{R}}_y + \omega^2\,{\tilde \mathcal{R}}_y+{\omega^2(\Qbar+\QPbar)-l(l+2)-3/4\over {\bar r}^2}\,{\tilde \mathcal{R}}_y+O({\bar r}^{-3})=0
\label{eqrsy}
\eea
In (\ref{rr}) and (\ref{eqrsy}) we have organized the powers of $1/{\bar r}$ by assuming that
$\tilde \mathcal{R}_s$, $\tilde \mathcal{R}_y$ and their ${\bar r}$-derivatives are of order ${\bar r}^0$: By looking at (\ref{eqrsy}), 
we see that this assumption is actually implied by the
 equations themselves. Note also that the next to leading corrections in
(\ref{eqrsy}) are of oder $1/{\bar r}^2$. We can thus conclude that $\mathcal{R}_s$ and $\mathcal{R}_y$ 
have the form
\bea
\mathcal{R}_s&\approx&{e^{i \omega\,{\bar r}}\over {\bar r}^{1/2}}\,\sum_{n=0}^\infty
{r_{s,+}^{(n)}\over {\bar r}^n}
+{e^{-i \omega\,{\bar r}}\over {\bar r}^{1/2}}\,\sum_{n=0}^\infty{r_{s,-}^{(n)}\over {\bar r}^n}\nonumber\\
\mathcal{R}_y&\approx&{e^{i \omega\,{\bar r}}\over {\bar r}^{3/2}}\,\sum_{n=0}^\infty{r_{y,+}^{(n)}\over {\bar r}^n}+
{e^{-i \omega\,{\bar r}}\over {\bar r}^{3/2}}\,\sum_{n=0}^\infty{r_{y,-}^{(n)}\over {\bar r}^n}
\eea
Analogously to the case of the scalar perturbation, the coefficients $r_{s,\pm}^{(n)}$ and $r_{y,\pm}^{(n)}$ for $n>1$ are recursively determined from
 $r_{s,\pm}^{(0)}$ and $r_{y,\pm}^{(0)}$, and are finite for any value of
$\omega$ and any finite $n$.
Substituting in (\ref{rr}) we have the solution for $\mathcal{R}_{\bar r}$
\be
\mathcal{R}_{\bar r}\approx{e^{i \omega\,{\bar r}}\over {\bar r}^{5/2}}\,\sum_{n=0}^\infty
{r_{r,+}^{(n)}\over {\bar r}^n}
+{e^{-i \omega\,{\bar r}}\over {\bar r}^{5/2}}\,\sum_{n=0}^\infty{r_{r,-}^{(n)}\over {\bar r}^n}
\label{rrfinal}
\ee 
where $r_{{\bar r},\pm}^{(n)}$ are determined in terms of $r_{s,\pm}^{(n)}$ and $r_{y,\pm}^{(n)}$ 
(the leading coefficients $r_{{\bar r},\pm}^{(0)}$ vanish for $l=0$).

The components $\mathcal{R}_s$ and $\mathcal{R}_y$ give rise to nonvanishing energy flux at 
infinity while $\mathcal{R}_{\bar r}$
does not contribute to the flux at leading order. Note that because
$\mathcal{R}_{\bar r}$ is zero if both $\mathcal{R}_s$ and $\mathcal{R}_y$ vanish, it is not possible to have a solution in which only $\mathcal{R}_{\bar r}$ is 
excited, and thus all solutions carry some flux at infinity.


\begin{thebibliography}{99}
\bibitem{vafa}
C.~Vafa,
Nucl.\ Phys.\ B {\bf 463}, 435 (1996)
[arXiv:hep-th/9512078].

\bibitem{sen}
A.~Sen,
Nucl.\ Phys.\ B {\bf 440}, 421 (1995)
[arXiv:hep-th/9411187];
A.~Sen,
Mod.\ Phys.\ Lett.\ A {\bf 10}, 2081 (1995)
[arXiv:hep-th/9504147].

\bibitem{lm4}
O.~Lunin and S.~D.~Mathur,
Nucl.\ Phys.\ B {\bf 623}, 342 (2002)
[arXiv:hep-th/0109154].

\bibitem{supertube}
D.~Mateos and P.~K.~Townsend,
 Phys.\ Rev.\ Lett.\  {\bf 87}, 011602 (2001)
[arXiv:hep-th/0103030];
R.~Emparan, D.~Mateos and P.~K.~Townsend,
JHEP {\bf 0107}, 011 (2001)
[arXiv:hep-th/0106012];
D.~Mateos, S.~Ng and P.~K.~Townsend,
JHEP {\bf 0203}, 016 (2002)
[arXiv:hep-th/0112054].
\bibitem{baka}
D.~s.~Bak and A.~Karch,
Nucl.\ Phys.\ B {\bf 626}, 165 (2002)
[arXiv:hep-th/0110039].
\bibitem{kmpw}
M.~Kruczenski, R.~C.~Myers, A.~W.~Peet and D.~J.~Winters,
JHEP {\bf 0205}, 017 (2002)
[arXiv:hep-th/0204103].

\bibitem{dkps}
 M.~R.~Douglas, D.~Kabat, P.~Pouliot and S.~H.~Shenker,
Nucl.\ Phys.\ B {\bf 485}, 85 (1997)
[arXiv:hep-th/9608024].
  
\bibitem{kami}
D.~M.~Kaplan and J.~Michelson,
Phys.\ Lett.\ B {\bf 410}, 125 (1997)
[arXiv:hep-th/9707021].

\bibitem{mist}
J.~Michelson and A.~Strominger,
JHEP {\bf 9909}, 005 (1999)
[arXiv:hep-th/9908044].

\bibitem{khuri}
R.~R.~Khuri,
Phys.\ Lett.\ B {\bf 294}, 331 (1992)
[arXiv:hep-th/9205052].

\bibitem{lm3}
O.~Lunin and S.~D.~Mathur,
Nucl.\ Phys.\ B {\bf 615}, 285 (2001)
[arXiv:hep-th/0107113].

\bibitem{lm6}
O.~Lunin, S.~D.~Mathur and A.~Saxena,
Nucl.\ Phys.\ B {\bf 655}, 185 (2003)
[arXiv:hep-th/0211292].


\bibitem{pm}
B.~C.~Palmer and D.~Marolf,
JHEP {\bf 0406}, 028 (2004)
[arXiv:hep-th/0403025].

\bibitem{emission}
S.~D.~Mathur,
Nucl.\ Phys.\ B {\bf 529}, 295 (1998)
[arXiv:hep-th/9706151].

\bibitem{review}
 S.~D.~Mathur,
arXiv:hep-th/0510180.

\bibitem{3charge}
S.~D.~Mathur, A.~Saxena and Y.~K.~Srivastava,
Nucl.\ Phys.\ B {\bf 680}, 415 (2004)
[arXiv:hep-th/0311092];
S.~Giusto, S.~D.~Mathur and A.~Saxena,
Nucl.\ Phys.\ B {\bf 701}, 357 (2004)
[arXiv:hep-th/0405017];
S.~Giusto, S.~D.~Mathur and A.~Saxena,
arXiv:hep-th/0406103;
O.~Lunin,
JHEP {\bf 0404}, 054 (2004)
[arXiv:hep-th/0404006].

\bibitem{gmr}
J.~P.~Gauntlett, J.~B.~Gutowski, C.~M.~Hull, S.~Pakis and H.~S.~Reall,
Class.\ Quant.\ Grav.\  {\bf 20}, 4587 (2003)
[arXiv:hep-th/0209114];
J.~B.~Gutowski, D.~Martelli and H.~S.~Reall,
Class.\ Quant.\ Grav.\  {\bf 20}, 5049 (2003)
[arXiv:hep-th/0306235].

\bibitem{marika}
M.~Taylor,
arXiv:hep-th/0507223.

\bibitem{lm2}
O.~Lunin and S.~D.~Mathur,
Nucl.\ Phys.\ B {\bf 610}, 49 (2001) 
[hep-th/0105136].

\bibitem{lm5}
O.~Lunin and S.~D.~Mathur,
Phys.\ Rev.\ Lett.\  {\bf 88}, 211303 (2002)
[arXiv:hep-th/0202072].

\bibitem{fuzz}
S.~D.~Mathur,
Fortsch.\ Phys.\  {\bf 53}, 793 (2005)
[arXiv:hep-th/0502050].

\bibitem{supercurve}
D.~Mateos, S.~Ng and P.~K.~Townsend,
Phys.\ Lett.\ B {\bf 538}, 366 (2002)
[arXiv:hep-th/0204062].

\bibitem{bho}
 D.~Bak, Y.~Hyakutake and N.~Ohta,
Nucl.\ Phys.\ B {\bf 696}, 251 (2004)
[arXiv:hep-th/0404104];
D.~Bak, Y.~Hyakutake, S.~Kim and N.~Ohta,
Nucl.\ Phys.\ B {\bf 712}, 115 (2005)
[arXiv:hep-th/0407253].

\bibitem{giant}
S.~R.~Das, A.~Jevicki and S.~D.~Mathur,
Phys.\ Rev.\ D {\bf 63}, 024013 (2001)
[arXiv:hep-th/0009019].

\bibitem{tachyon}
O.~Lunin, S.~D.~Mathur, I.~Y.~Park and A.~Saxena,
Nucl.\ Phys.\ B {\bf 679}, 299 (2004)
[arXiv:hep-th/0304007].

\bibitem{gv}
Vachaspati and T.~Vachaspati, 
Phys. \ Lett. \ B {\bf 238}, 41 (1990);
D.~Garfinkle and T.~Vachaspati,
Phys.\ Rev.\ D {\bf 42}, 1960 (1990);
A.~Dabholkar, J.~P.~Gauntlett, J.~A.~Harvey and D.~Waldram,
Nucl.\ Phys.\ B {\bf 474}, 85 (1996)
[arXiv:hep-th/9511053];
C.~G.~.~Callan, J.~M.~Maldacena and A.~W.~Peet,
Nucl.\ Phys.\ B {\bf 475}, 645 (1996)
[arXiv:hep-th/9510134].

\bibitem{lmm}
O.~Lunin, J.~Maldacena and L.~Maoz,
arXiv:hep-th/0212210.

\bibitem{ms}
J.~Maharana and J.~H.~Schwarz,
Nucl.\ Phys.\ B {\bf 390}, 3 (1993)
[arXiv:hep-th/9207016].

\bibitem{book}
A.~Erd\'elyi, ''Asymptotic Expansions,'' New York: Dover (1987).

\bibitem{reedsimon}
M.~Reed and B.~Simon,
``Methods of Modern Mathematical Physics. Vol. 4: Analysis of Operators,''
New York: Academic Press (1978); see Theorems XIII.56, XIII.57, XIII.58.

\bibitem{bal}
V.~Balasubramanian, J.~de Boer, E.~Keski-Vakkuri and S.~F.~Ross,
Phys.\ Rev.\ D {\bf 64}, 064011 (2001)
[arXiv:hep-th/0011217].

\bibitem{mm}
J.~M.~Maldacena and L.~Maoz,
JHEP {\bf 0212}, 055 (2002)
[arXiv:hep-th/0012025].

\bibitem{maldastrom}
J.~M.~Maldacena and A.~Strominger,
Phys.\ Rev.\ D {\bf 55}, 861 (1997)
[arXiv:hep-th/9609026].

\bibitem{callanmalda}
C.~G.~.~Callan and J.~M.~Maldacena,
Nucl.\ Phys.\ B {\bf 472}, 591 (1996)
[arXiv:hep-th/9602043].

\bibitem{benawarner}
I.~Bena and N.~P.~Warner,
arXiv:hep-th/0408106.

\bibitem{gauntlett}
J.~P.~Gauntlett and J.~B.~Gutowski,
Phys.\ Rev.\ D {\bf 71}, 045002 (2005)
[arXiv:hep-th/0408122].

\bibitem{sv}
A.~Strominger and C.~Vafa,
Phys.\ Lett.\ B {\bf 379}, 99 (1996)
[arXiv:hep-th/9601029].

\bibitem{gimon}
P.~Berglund, E.~G.~Gimon and T.~S.~Levi,
arXiv:hep-th/0505167.

\bibitem{bw2}
I.~Bena and N.~P.~Warner,
arXiv:hep-th/0505166.

\bibitem{kk}
I.~Bena and P.~Kraus,
Phys.\ Rev.\ D {\bf 72}, 025007 (2005)
[arXiv:hep-th/0503053];
D.~Gaiotto, A.~Strominger and X.~Yin,
arXiv:hep-th/0503217;
A.~Saxena, G.~Potvin, S.~Giusto and A.~W.~Peet,
arXiv:hep-th/0509214.

\bibitem{bekr}
I.~Bena and P.~Kraus,
Phys.\ Rev.\ D {\bf 70}, 046003 (2004)
[arXiv:hep-th/0402144].
\bibitem{bena}
I.~Bena,
Phys.\ Rev.\ D {\bf 70}, 105018 (2004)
[arXiv:hep-th/0404073].


\end{thebibliography}
\end{document}